\documentclass[twocolumn]{aastex631}  

\newcommand\data{\boldsymbol{v}}
\newcommand\datascalar{v}

\newcommand\gains{\boldsymbol{g}}
\newcommand\gain{g}

\newcommand\modelvals{\boldsymbol{m}}
\newcommand\modelval{m}

\newcommand\negloglikelihood{\chi^2}

\newcommand\radweight{W}
\newcommand\weightmat{\boldsymbol{\mathsf{W}}}

\newcommand\costcomp{r}

\usepackage{commath, graphicx, bbold, endnotes, graphicx, subfigure, multirow, afterpage}

\received{August 8, 2022}
\revised{December 5, 2022}
\accepted{December 17, 2022}
\published{February 1, 2023}

\begin{document}

\title{Delay-Weighted Calibration: Precision Calibration for 21 cm Cosmology with Resilience to Sky Model Error}

\author[0000-0003-4980-2736]{Ruby Byrne}
\affiliation{Astronomy Department, California Institute of Technology, 1200 E California Boulevard, Pasadena, CA, 91125, USA}



\begin{abstract}

One of the principal challenges of 21 cm cosmology experiments is overcoming calibration error. Established calibration approaches in the field require an exquisitely accurate sky model, and low-level sky model errors introduce calibration errors that corrupt the cosmological signal. We present a novel calibration approach called delay-weighted calibration, or DWCal, which enables precise calibration even in the presence of sky model error. Sky model error does not affect all power spectrum modes equally, and DWCal fits calibration solutions preferentially from error-free modes. We apply this technique to simulated data, showing that it substantially reduces calibration error in the presence of realistic levels of sky model error and can improve 21 cm power spectrum sensitivity by approximately 2 orders of magnitude.

\end{abstract}



\section{Introduction}
\label{s:intro}

Interferometric measurement of 21 cm emission from neutral hydrogen at high redshift has great potential for mapping large volumes of the universe, probing the Cosmic Dawn and Epoch of Reionization (EoR), and measuring large-scale structure evolution throughout cosmological history (see \citealt{Furlanetto2006}, \citealt{Pritchard2008}, \citealt{Morales2010}, and \citealt{Liu2020} for reviews). The success of 21 cm cosmology experiments is currently limited by the precision with which bright intervening emission, or foregrounds, can be separated from the faint cosmological signal. While the foreground emission is a staggering 4-5 orders of magnitude brighter than the 21 cm signal, it is very spectrally smooth and therefore can, in principle, be distinguished from the spectrally structured cosmological signal.

However, the success of this approach requires extremely precise frequency-dependent, or bandpass, calibration. Low-level spectral errors in calibration introduce spectral structure into the foreground signal, preventing effective foreground removal. In recent years, this problem has inspired a proliferation of precision bandpass calibration methods \citep{Mitchell2008, Yatawatta2009, Kazemi2011, Sullivan2012, Kazemi2013, Salvini2014, Sievers2017, Dillon2020, Sob2020, Kern2020, Byrne2021b, ewall-wice2022, Sims2022, Sims2022b}. Nevertheless, precision calibration approaches generally require exquisite prior knowledge of the sky signal and instrumental response, and calibration precision remains a principal limitation of 21 cm cosmology analyses.

We introduce a novel approach to bandpass calibration called delay-weighted calibration, or DWCal, which combats spectral calibration error. DWCal imposes no prior assumptions about the instrument's bandpass response and can accurately calibrate even when the instrumental bandpass has substantial frequency structure. It fits the same number of free calibration parameters as traditional sky-based calibration, avoiding the computational challenges and overfitting concerns associated with many-parameter calibration approaches. DWCal can be combined with other advanced calibration techniques such as redundant calibration or unified calibration.

DWCal specifically combats a mechanism of calibration error by which errors in the calibration sky model introduce spurious spectral structure into the calibration solutions. Missing or mismodeled sources and structures in the calibration sky model produce bandpass calibration errors, even when both the true sky signal and the sky model are intrinsically spectrally smooth. For low levels of sky model error, the resulting calibration error is nonetheless at a level that precludes a detection of the cosmological 21 cm signal \citep{Barry2016, Trott2016, Ewall-Wice2017, Byrne2019}. 

DWCal joins a suite of techniques designed to combat this error mechanism. First is simply the development of highly accurate and complete sky models \citep{Yatawatta2013a, Offringa2015, Carroll2016, Hurley-Walker2017, Patil2017, Eastwood2018, deGasperin2020, Byrne2021c}. These have substantially improved calibration performance, but some degree of sky model error is nonetheless unavoidable. Next, assuming stable and uniform antenna responses, calibration benefits from averaging across antennas and time \citep{Barry2016, Kern2020}. If the intrinsic antenna response is relatively spectrally smooth, the calibration solutions can be averaged across frequency or fit to a simple function such as a low-order polynomial \citep{Yatawatta2015, Barry2016, Dillon2018, Gehlot2018, Barry2019b, Li2019, Eastwood2019, Mertens2020}. Short baselines produce less spectral contamination in calibration, so calibration to short baselines only can improve calibration performance \citep{Ewall-Wice2017, Patil2017}. Taking this idea to its extreme, \citealt{Barry2019b} and \citealt{Li2019} calibrate with baselines of zero length, using the autocorrelation visibilities to fit the antennas' bandpass responses.

Other calibration approaches address this error mechanism by incorporating additional calibration parameters to directly fit the sky signal. Redundant calibration, applicable to highly regular arrays, constrains parameters of the sky signal by matching repeated baseline measurements, thereby reducing reliance on the sky model \citep{Wieringa1992, Liu2010, Grobler2018, Dillon2018, Li2018, Dillon2020, Kern2020, Zhang2020}. However, regular arrays are particularly sensitive to sky model error, and they experience substantial bandpass calibration error from low-level sky model error even when redundantly calibrated \citep{Byrne2019}. \citealt{ewall-wice2022} directly fits the spectrally smooth components of the sky signal measured by each baseline, fitting out much of the sky model error at the cost of a large increase in calibration parameters.

The DWCal framework takes a new approach to tackling the problem of sky model error in calibration. Interferometric measurements reconstruct spectrally smooth signals in a compact region of 2D power spectrum space (here the two dimensions refer to power spectrum modes parallel and perpendicular to the line of sight, respectively). This region is known as the ``foreground wedge'' and has been explored in depth in the literature \citep{Morales2012, Trott2012, Vedantham2012, Pober2013, Thyagarajan2013, Hazelton2013, Dillon2015, Morales2019}. Consequently, sky model error appears within this foreground wedge region. DWCal exploits this feature, incorporating knowledge of the foreground wedge to downweight contaminated power spectrum modes in calibration.

Unlike short baseline calibration or autocorrelation calibration, DWCal calibrates with all baseline measurements, employing the full measurement set to constrain the calibration solutions. However, each baseline preferentially constrains the modes on which it is unaffected by sky model error. It assumes that the sky model error is compact in the 2D power spectrum space but makes no such assumption of the instrumental response, so it can fit any instrumental spectral features. It does not require averaging across antennas or time and supports arrays with heterogeneous antenna responses and time dependence. DWCal does not introduce any additional calibration degeneracies and has the same number of degenerate parameters as sky-based calibration (see \S\ref{s:regularization} for a discussion of these degeneracies). While it is not immune to sky model error --- DWCal still requires a good sky model --- we show that it is significantly more robust against sky model error than established precision calibration techniques.

DWCal introduces cross-frequency constraints on the gains. This means that like calibration approaches such as those discussed in \citealt{Yatawatta2015}, \citealt{Mertens2020}, \citealt{ewall-wice2022}, and \citealt{Sims2022}, DWCal cannot be parallelized across frequency. Instead, the gains at all frequencies must be jointly calculated.

This paper introduces a simple implementation of DWCal, demonstrating the principles of the technique. This implementation can be considered an extension to simple sky-based calibration. We apply DWCal to simulated data, demonstrating that it improves calibration performance in the presence of sky model error. Future work could optimize the DWCal algorithm to reduce its computational cost. It could also combine DWCal with other calibration techniques such as redundant calibration or unified calibration.

\section{DWCal Formalism}

In general, interferometric calibration consists of minimizing a cost function, often denoted $\chi^2$, to constrain parameters of the instrument response. The fundamental innovation underlying DWCal is the incorporation of delay-dependent expected error into the calibration $\chi^2$. In \S\ref{s:basic_formalism} we derive the DWCal $\chi^2$ as an extension to simple sky-based calibration. In \S\ref{s:regularization} we present a new approach to constraining calibration degeneracies, and in \S\ref{s:extensions} we discuss further extensions to DWCal.

\subsection{The Basic DWCal Formalism}
\label{s:basic_formalism}

In its simplest form, sky-based calibration consists of calculating gain values that minimize the quantity
\begin{equation}
    \negloglikelihood = \sum_f \sum_{jk} \left| \gain_j(f) \gain_k^*(f) \datascalar_{jk}(f) - \modelval_{jk}(f) \right|^2.
\label{eq:sky_cal_simple}
\end{equation}
Here $f$ represents frequencies and $j$ and $k$ index antennas. $\sum_f$ indicates the sum over all frequency channels and $\sum_{jk}$ indicates the sum across all baselines; $\datascalar_{jk}(f)$ is the visibility derived by correlating antennas $j$ and $k$ at frequency $f$; $\modelval_{jk}(f)$ is a model of that visibility derived from a sky model propagated through an instrument simulator; $\gains(f)$ are the antenna gains, parameterized as a complex number for each antenna and frequency, and the asterisk $*$ denotes the complex conjugate. We do not consider time dependence in this expression and assume a single time step. We further assume a single instrumental polarization mode.

Note that here we define a calibration convention in which the gains multiply the data, not the model visibilities. This diverges from the convention typically used in the literature (see, for example, \citealt{Hamaker1996a}) for reasons delineated below. Under our convention, we calibrate data by multiplying the raw visibilities by the calculated gains.

Minimizing Equation \ref{eq:sky_cal_simple} produces a maximum-likelihood estimate of the gains, assuming independent Gaussian-distributed thermal noise and model visibility error at each frequency and baseline (see \citealt{Byrne2021b} for a derivation of this expression based on the Bayesian likelihood function). Because Equation \ref{eq:sky_cal_simple} assumes uncorrelated frequency channels, it is separable in frequency. Each frequency can be calibrated independently.

This simple sky-based calibration could be augmented with a frequency- and baseline-dependent weighting function. For example, we could replace Equation \ref{eq:sky_cal_simple} with an expression
\begin{equation}
    \negloglikelihood = \sum_f \sum_{jk} \radweight_{jk}(f) \left| \gain_j(f) \gain_k^*(f) \datascalar_{jk}(f) - \modelval_{jk}(f) \right|^2,
\label{eq:sky_cal_weighted}
\end{equation}
where $\radweight_{jk}(f)$ is a positive, scalar weighting function. $\radweight_{jk}(f)$ can be adjusted to downweight particular baselines and frequency channels with known contamination. For example, if we knew the model visibility $m_{jk}(f)$ was particularly inaccurate at frequency $f = f_0$, we could reduce the value of $\radweight_{jk}(f_0)$ to protect the calibration solutions from that error.

Calibrating with Equation \ref{eq:sky_cal_weighted} is appropriate when the model visibility error appears in distinct frequency channels. However, sky model error produces error in the model visibilities at all frequencies. This model visibility error is not compact in the frequency domain but \textit{is} compact in the Fourier dual, or delay domain \citep{Parsons2009, Parsons2012, Pober2013, Morales2019}. We therefore aim to remap the calibration problem into delay space.

We return to Equation \ref{eq:sky_cal_simple}. To make our expressions more concise, we define a new quantity
\begin{equation}
    \costcomp_{jk}(f) = \gain_j(f) \gain_k^*(f) \datascalar_{jk}(f) - \modelval_{jk}(f).
\end{equation}
Equation \ref{eq:sky_cal_simple} can then be written as
\begin{equation}
    \negloglikelihood = \sum_{jk} \sum_f \left| \costcomp_{jk}(f) \right|^2.
\end{equation}
From the Plancherel theorem, this is equivalent to
\begin{equation}
    \negloglikelihood = \frac{1}{N_\text{freq}} \sum_{jk} \sum_{\eta} \, \left| \widetilde{\costcomp}_{jk}(\eta) \right|^2,
\label{eq:delay_space_chisquared}
\end{equation}
where $N_\text{freq}$ is the number of frequency channels and $\eta$ is delay, the Fourier dual of frequency $f$ with units of time. The tilde denotes the Fourier transformed quantity:
\begin{equation}
    \widetilde{\costcomp}_{jk}(\eta) = \sum_f \costcomp_{jk}(f) e^{-2 \pi i \eta f}.
\end{equation}
Via the convolution theorem
\begin{equation}
    \widetilde{\costcomp}_{jk}(\eta) = \widetilde{\gain}_j(\eta) * \widetilde{\gain}_k^*(-\eta) * \widetilde{\datascalar}_{jk}(\eta) - \widetilde{\modelval}_{jk}(\eta),
\label{eq:convolution_expression}
\end{equation}
where $*$ indicates the convolution.

Equation \ref{eq:delay_space_chisquared} represents a delay-space reformulation of the calibration problem. We can now introduce a delay-dependent weighting function, such that
\begin{equation}
    \negloglikelihood = \frac{1}{N_\text{freq}} \sum_{jk} \sum_{\eta} \widetilde{\radweight}_{jk}(\eta) \left| \widetilde{\costcomp}_{jk}(\eta) \right|^2.
\label{eq:delay_weighting}
\end{equation}
$\widetilde{\radweight}_{jk}(\eta)$ can now be tuned to downweight particular delay modes for a given baseline $\{ j, k \}$. This allows us to capture the delay dependence of the model visibility error. At this point we leave the weighting function $\widetilde{\radweight}_{jk}(\eta)$ fully general. In the next section we describe development and implementation of a specific weighting function targeting model visibility error in the foreground wedge.

This treatment motivates the calibration convention defined above, in which the gains multiply the data. If we were to use the convention established in \citealt{Hamaker1996a}, the model visibilities in Equation \ref{eq:convolution_expression} would be convolved with the gains. Nonzero gain values at high delay would spread model error beyond the foreground wedge modes, meaning that model error would no longer inhabit a compact delay-space region.

Equation \ref{eq:delay_weighting} achieves the delay-dependent weighting we desire. However, the double convolution in Equation \ref{eq:convolution_expression} makes this an unweildy expression to evaluate directly. We therefore transform back into the frequency domain.

Expanding Equation \ref{eq:delay_weighting} gives
\begin{equation}
    \negloglikelihood = \frac{1}{N_\text{freq}} \sum_{jk} \sum_\eta \widetilde{\radweight}_{jk}(\eta) \bigg| \sum_f \costcomp_{jk}(f) e^{-2 \pi i \eta f} \bigg|^2,
\end{equation}
or
\begin{equation}
    \negloglikelihood = \frac{1}{N_\text{freq}} \sum_{jk} \sum_\eta \widetilde{\radweight}_{jk}(\eta) \sum_f \sum_{f'} \costcomp_{jk}(f) \costcomp^*_{jk}(f') e^{2 \pi i \eta (f'-f)}.
\end{equation}
We can rewrite this expression as
\begin{equation}
    \negloglikelihood = \sum_{jk} \sum_f \sum_{f'} \bigg[ \frac{1}{N_\text{freq}} \sum_\eta \widetilde{\radweight}_{jk}(\eta) e^{2 \pi i \eta (f'-f)} \bigg] \costcomp_{jk}(f) \costcomp^*_{jk}(f'),
\end{equation}
where the bracketed expression is simply the inverse Fourier transform of the weighting function. We then get that
\begin{equation}
    \negloglikelihood = \sum_{jk} \sum_f \sum_{f'} \radweight_{jk}(f'-f) \costcomp_{jk}(f) \costcomp^*_{jk}(f').
\end{equation}
Expanding the quantity $\costcomp_{jk}(f)$ gives
\begin{equation}
\begin{split}
    \negloglikelihood = & \sum_{jk} \sum_f \sum_{f'} \radweight_{jk}(f'-f) \\
    & \times \left[\gain_j(f) \gain_k^*(f) \datascalar_{jk}(f) - \modelval_{jk}(f) \right] \\
    & \times \left[\gain_j(f') \gain_k^*(f') \datascalar_{jk}(f') - \modelval_{jk}(f') \right]^*.
\end{split}
\label{eq:dwcal}
\end{equation}
Note that this can be written more concisely as a matrix multiplication operation:
\begin{equation}
    \negloglikelihood = \sum_{jk} \left[\gains_j \gains_k^* \data_{jk} - \modelvals_{jk} \right]^\dag \weightmat_{jk} \left[\gains_j \gains_k^* \data_{jk} - \modelvals_{jk} \right],
\label{eq:dwcal_matrix}
\end{equation}
where $\gains_j$, $\data_{jk}$, and $\modelvals_{jk}$ are each vectors of length $N_\text{freq}$. $\weightmat_{jk}$ is an $N_\text{freq} \times N_\text{freq}$ matrix, and $\dag$ denotes the conjugate transpose.

Equations \ref{eq:dwcal} and \ref{eq:dwcal_matrix} take the form of a maximum-likelihood estimate with correlated frequency channels. We find that if model error is compact in delay it is necessarily not independent across frequencies. The function $\radweight_{jk}(f'-f)$ can be interpreted as encoding covariances between frequency channels.

If the weighting function is appropriately set, it reduces calibration's dependence on error-prone delay modes such as the foreground wedge modes. We recover sky-based calibration in the limit that $\radweight_{jk}(f'-f) = 0$ when $f \ne f'$.

\subsection{Constraining the Degenerate Phase}
\label{s:regularization}

Like sky-based calibration, DWCal is degenerate in the overall complex phase of the gains at each frequency. From Equation \ref{eq:dwcal}, note that for each frequency $f$ the transformation $\gains(f) \rightarrow \gains(f) e^{i \phi}$ leaves $\negloglikelihood$ unchanged, where $\phi$ is an arbitrary real constant.

This degeneracy is typically constrained by setting the phase of a reference antenna's gain to zero or requiring that the average phase of the gains across all antennas be zero. This constraint can be imposed after minimizing the degenerate $\negloglikelihood$. However, this risks degrading the performance of an iterative optimization algorithm as it cannot fit to a unique minimum. We choose instead to introduce a regularization term to break the phase degeneracy.

We use L2 regularization to constrain the mean phase at each frequency to be zero. Equation \ref{eq:dwcal} then becomes
\begin{equation}
\begin{split}
    \negloglikelihood = & \sum_{jk} \sum_f \sum_{f'} \radweight_{jk}(f'-f) \\
    & \times \left[\gain_j(f) \gain_k^*(f) \datascalar_{jk}(f) - \modelval_{jk}(f) \right] \\
    & \times \left[\gain_j(f') \gain_k^*(f') \datascalar_{jk}(f') - \modelval_{jk}(f') \right]^* \\
    + & \lambda \sum_f \bigg( \sum_j \operatorname{Arg} \big[ \gain_j(f) \big] \bigg)^2,
\end{split}
\label{eq:dwcal_regularized}
\end{equation}
where the final term imposes the phase regularization. Here $\operatorname{Arg}$ denotes the complex phase and $\sum_j$ indicates the sum over all antennas; $\lambda$ is an arbitrary positive, real constant. In principle the value of $\lambda$ does not affect the global minimum of $\negloglikelihood$, as the global minimum occurs where the regularization term is identically zero. However, in practice we found that tuning $\lambda$ impacts the speed and accuracy of the optimization algorithm.

Equation \ref{eq:dwcal_regularized} produces unique, nondegenerate solutions, and we use this expression to calculate the calibration solutions presented throughout this paper. Because it constrains the mean phase to be zero, we cannot calibrate a nonzero absolute phase at any frequency.

\subsection{Further Extensions to DWCal}
\label{s:extensions}

The DWCal formalism described above and represented by Equations \ref{eq:dwcal} and \ref{eq:dwcal_regularized} can be considered an extension to traditional sky-based calibration (Equation \ref{eq:sky_cal_simple}). However, the principle behind DWCal can be applied to other, more advanced calibration techniques.

DWCal can be applied to redundant calibration, specifically to the absolute calibration step that uses a sky model to constrain the bulk array response \citep{Liu2010, Dillon2018, Li2018, Kern2020}. Absolute calibration is susceptible to spectral error from low-level errors in the sky model \citep{Byrne2019}. Using the DWCal formalism to downweight the foreground wedge modes in absolute calibration could reduce the impact of sky model error on redundant calibration solutions.

Unified calibration represents a middle ground between sky-based and redundant calibration \citep{Byrne2021b}. It directly fits the sky signal measured by each baseline or redundant baseline set, and it uses a sky model to impose a prior on that fit. DWCal can be applied to unified calibration by incorporating a delay-dependent weighting function into the prior. This would decrease the strength of the prior in foreground wedge modes and improve unified calibration's spectral performance.

Throughout this section we have assumed calibration of a single polarization. However, DWCal can be extended to fully polarization calibration. For a dual-polarization interferometer, fully polarized calibration parameterizes the gains as a $2\times2$ polarization matrix for each antenna and frequency interval \citep{Sault1996}. If an instrument experiences negligible cross-polarization signal coupling, the gains can be more simply parameterized with two values per antenna and frequency, corresponding to the two instrumental polarizations \citep{Byrne2022a}. DWCal is fully applicable to polarized calibration, and the delay-dependent weighting function can be applied to each polarization mode.

Gains need not be parameterized per frequency. While it is common practice to fit parameters of a low-order polynomial or other simple function to the calibrated bandpass after calculating per-frequency gains \citep{Barry2016, Dillon2018, Gehlot2018, Barry2019b, Li2019, Eastwood2019, Mertens2020}, these parameters can instead be calculated directly during the $\chi^2$ minimization operation \citep{Byrne2021b}. This technique could be applied to DWCal, in which case the gains in Equations \ref{eq:dwcal} or \ref{eq:dwcal_regularized} would be considered functions of the tunable calibration parameters. However, it is critical that the gain parameterization appropriately represents the true instrumental response in order for calibration to accurately capture all spectral features. Inaccurate gain parameterizations risk introducing additional calibration error.

For the remainder of this paper we focus on the simplest version of DWCal and calibrate by minimizing Equation \ref{eq:dwcal_regularized}. However, we emphasize that DWCal is not an alternative to precision calibration approaches such as redundant or unified calibration. Rather, the DWCal technique can supplement established calibration approaches to reduce spectral calibration error and build resilience to sky model error.

\section{Methods}

We demonstrate the DWCal technique with simulated data and a simple off-the-shelf optimization algorithm. The calibration code is written in Python and is publicly available on GitHub\footnote{\texttt{https://github.com/rlbyrne/dwcal}} under a BSD 3-Clause License as well as archived in Zenodo \citep{DWCal2022}. We use \textsc{pyuvdata}\footnote{\texttt{https://github.com/RadioAstronomySoftwareGroup/pyuvdata}} \citep{Hazelton2017}  for interfacing with the visibility data and calibration solutions.

\subsection{Visibility Simulation}
\label{s:vis_sim}

We simulate visibilities with \textsc{fhd}\footnote{\texttt{https://github.com/EoRImaging/FHD}} \citep{Sullivan2012, Barry2019a, Byrne2022a, Sullivan2022}. The simulation is based on the Murchison Widefield Array (MWA) Phase I configuration \citep{Tingay2013}. This array consists of 127 antennas in a nonregular imaging configuration. The \textsc{fhd} simulation corresponds to a zenith-pointed observation of the MWA's ``EoR-0'' field, centered at R.A.\ $0^\text{h}$, decl.\ $-27^\circ$. The simulation spans a frequency range of 167-198 MHz at a frequency resolution of 80 kHz. It corresponds to a 2 minute snapshot image with 2 s time resolution; for calibration, we average the visibilities across the full 2 minutes. We use the beam model developed by \citealt{Sutinjo2015} and average it across the full frequency range to represent a frequency-invariant beam. We simulate only a single polarization, corresponding to the east--west aligned dipoles.

We represent the sky signal with the GLEAM source catalog \citep{Hurley-Walker2017}, and we model point sources only. The simulated ``true'' visibilities are derived from the full catalog. Following the approach used in \citealt{Barry2016} and \citealt{Byrne2019}, we simulate model visibilities $\modelvals(f)$ from an incomplete catalog that omits the faintest GLEAM sources. Our incomplete catalog includes 90\% of the full catalog's power and omits sources fainter than 91 mJy.

\begin{figure}
    \centering
    \includegraphics[width=\columnwidth]{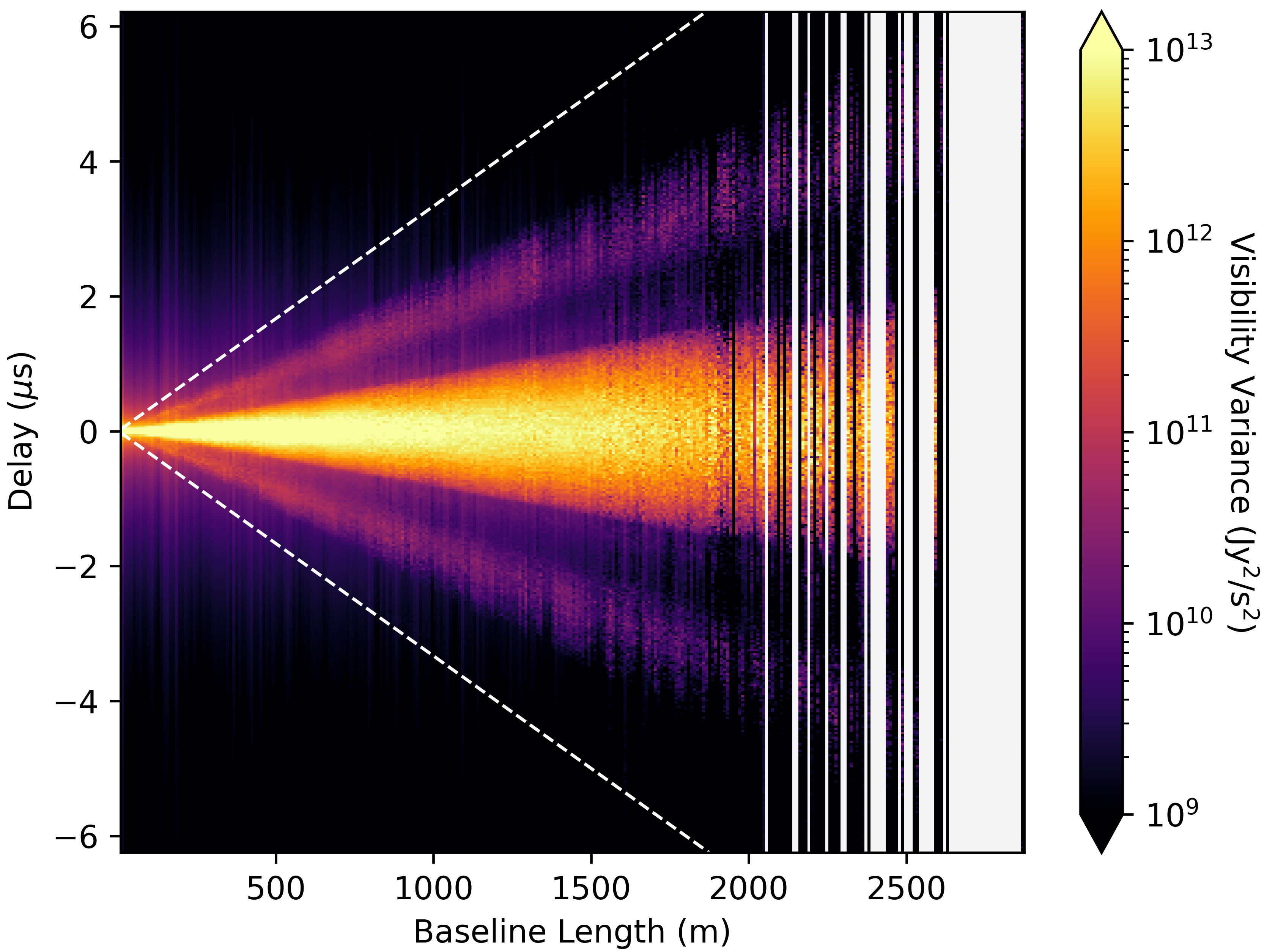}
    \caption{Plot of simulated model visibility error as a function of delay mode and baseline length. The visibility error is calculated by differencing the simulated true visibilities and model visibilities. The differenced visibilities are then Fourier transformed across frequency and binned by baseline length, and we plot the variance in each bin. Model visibility error results from faint missing sources in the simulated sky model, and, due to the foreground wedge effect, these errors produce a wedge-shaped feature in delay space. The affected delay modes depend on the sources' zenith angles, with sources on the horizon producing error out to the dashed white lines. The foreground wedge seen here does not extend to the horizon line due to the finite extent of the field of view. Missing data reflects sparse baseline coverage at long baselines.}
    \label{fig:model_error}
\end{figure}

Figure \ref{fig:model_error} plots model visibility error as a function of delay mode and baseline length. Due to the foreground wedge effect, the missing sources in the sky model produce model visibility error predominantly in a wedge-shaped region of delay space.

We simulate the instrument response by defining randomized gains for each antenna and frequency. We then divide the true visibilities simulated from the full GLEAM catalog by the gains to derive uncalibrated data visiblities $\data(f)$. Calibration then attempts to recover the gains as accurately as possible, enabling reconstruction of the simulated true signal from the data visibilities. Our simulation does not include thermal noise, so any calibration error results from sky model error only.

\subsection{Defining the Weighting Function}
\label{s:weighting_func}

DWCal uses a weighting function that represents the model visibility error for each baseline and delay mode. The form of this weighting function is fully arbitrary, but calibration performance improves with weightings that accurately estimate the true model visibility error.

\begin{figure}
    \centering
    \includegraphics[width=\columnwidth]{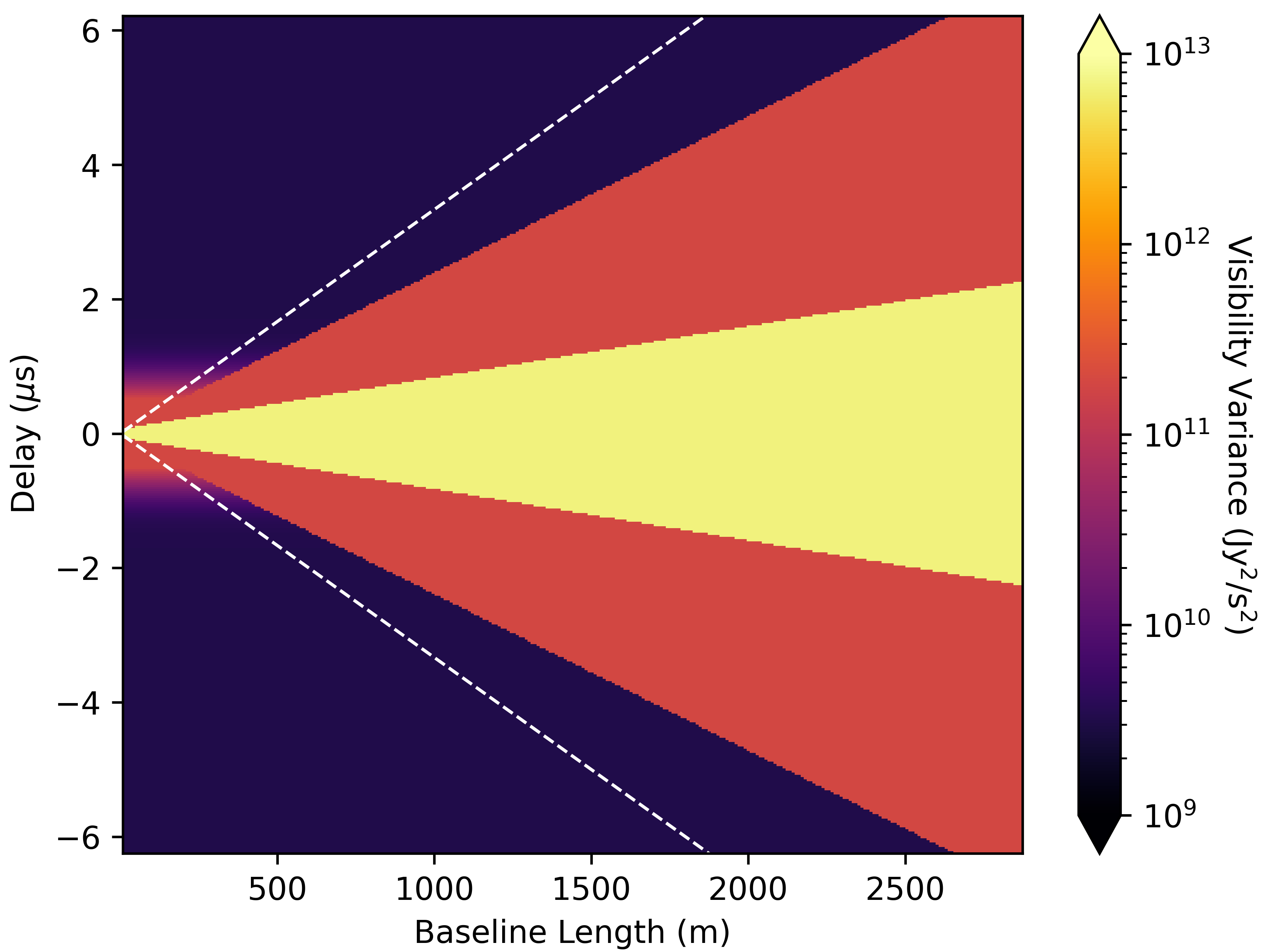}
    \caption{Plot of the estimated model visibility error used in calibration. We choose to represent the model visibility error, plotted in Figure \ref{fig:model_error}, with the values plotted here. We divide the foreground wedge region into two parts, corresponding to an inner and outer region, and calculate the average model visibility error in each region. Furthermore, for short baselines we model leakage near delays $\eta \approx 0$ with a decaying exponential function. In calibration we set the weighting function $\widetilde{\radweight}_{jk}(\eta)$ (see Equation \ref{eq:delay_weighting}) equal to the reciprocal of these values. As in Figure \ref{fig:model_error}, the dashed white lines denote the horizon extent.}
    \label{fig:weighting_function}
\end{figure}

We choose to define the weighting function with a relatively simple parameterization of the model visibility error, presented in Figure \ref{fig:weighting_function}. We divide the foreground wedge into inner and outer wedge regions and empirically calculate the variance in each region from the model visibility error (plotted in Figure \ref{fig:model_error}). We further represent short baseline error near delays $\eta \approx 0$ by fitting a decaying exponential function in $|\eta|$.

From Equation \ref{eq:delay_weighting}, $\widetilde{\radweight}_{jk}(\eta)$ can be interpreted as the reciprocal of the variance of the error on models of baseline $\{j, k\}$ at delay mode $\eta$. We therefore set $\widetilde{\radweight}_{jk}(\eta)$ equal to the reciprocal of the values plotted in Figure \ref{fig:weighting_function}.

\begin{figure*}
    \centering
    \includegraphics[width=2\columnwidth]{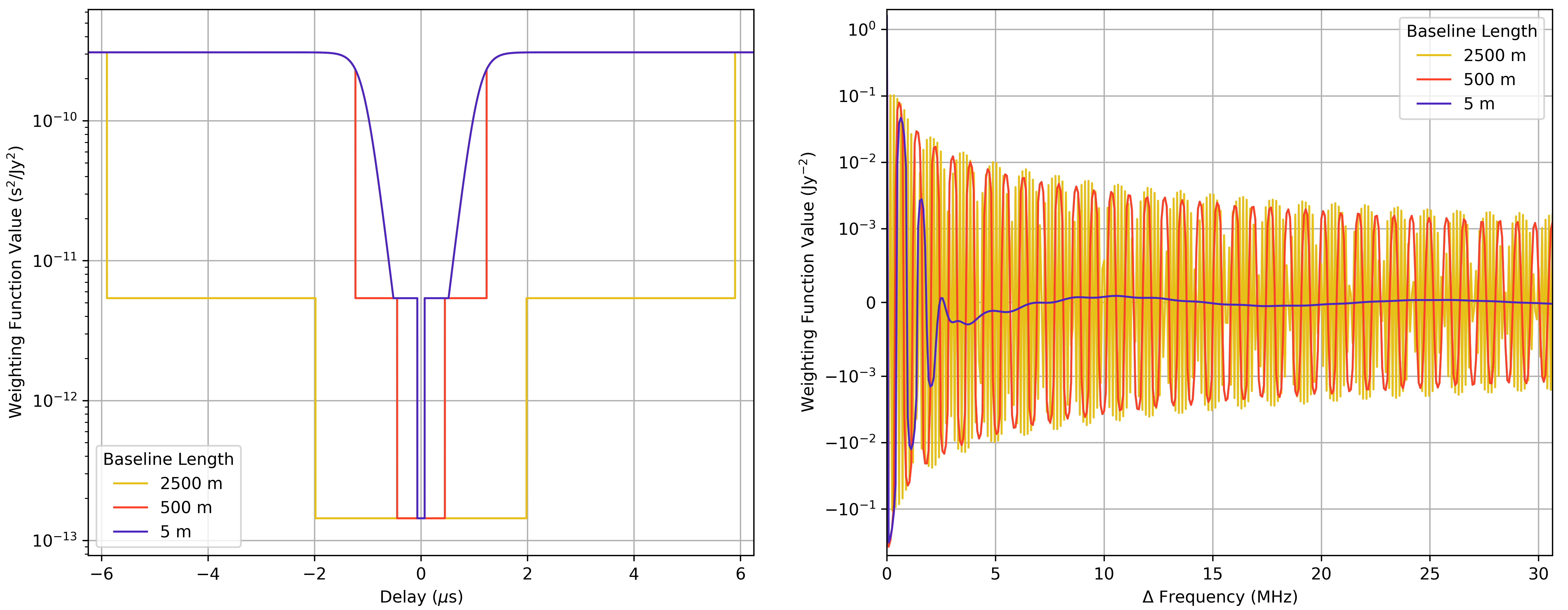}
    \caption{Plots of the weighting function for three baseline lengths. On the left are examples of the weighting function $\widetilde{\radweight}_{jk}(\eta)$, plotted as a function of delay $\eta$. $\widetilde{\radweight}_{jk}(\eta)$ is defined as the reciprocal of the estimated model visibility error variance, plotted in Figure \ref{fig:weighting_function}. On the right are the inverse Fourier transformed quantities $\radweight_{jk}(\Delta f)$.}
    \label{fig:weighting_func_select_bls}
\end{figure*}

\begin{figure*}
    \centering
    \includegraphics[width=2\columnwidth]{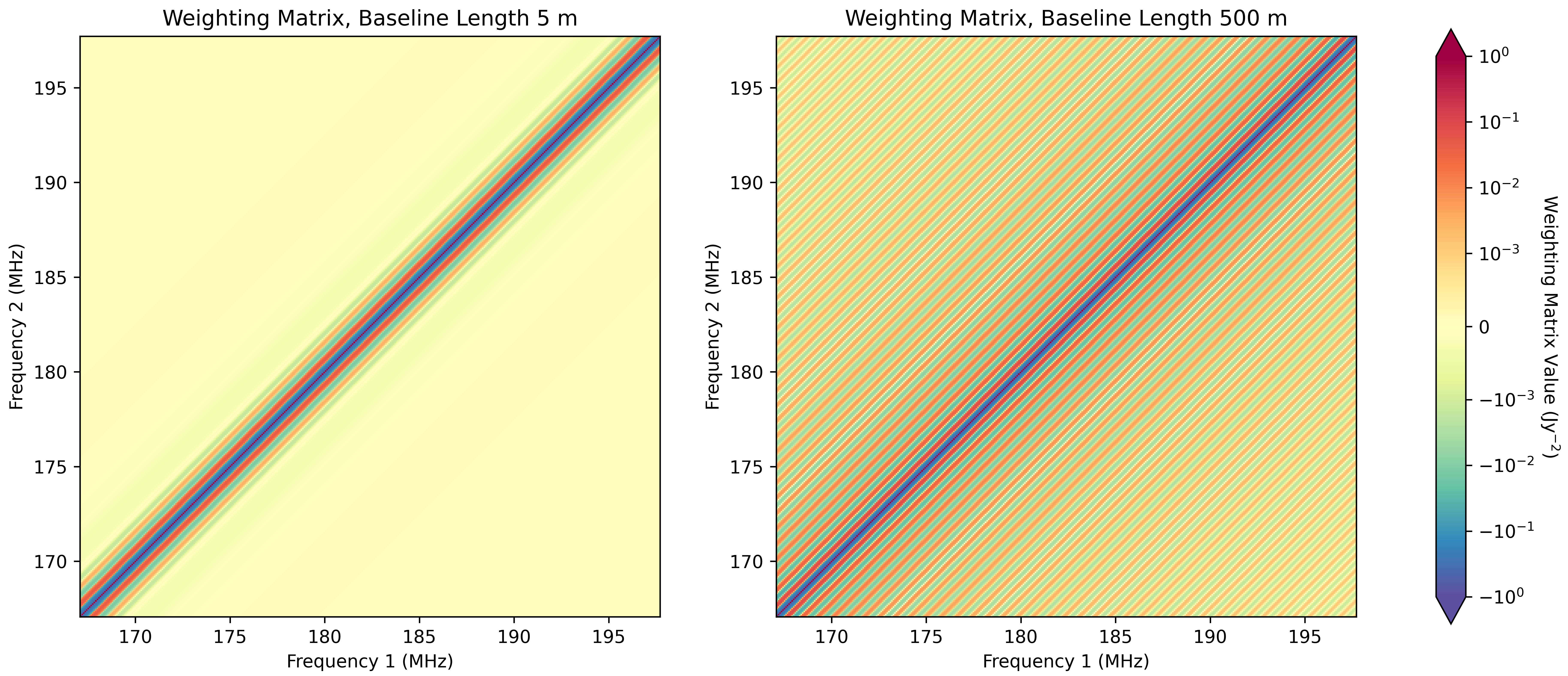}
    \caption{Examples of weighting matrices $\weightmat_{jk}$ used in calibration, for baselines of length 5 m (left) and 500 m (right). The weighting functions plotted in Figure \ref{fig:weighting_func_select_bls} are transformed into these $N_\text{freq} \times N_\text{freq}$ matrices for evaluating $\chi^2$ in calibration.}
    \label{fig:weighting_matrices}
\end{figure*}

Figure \ref{fig:weighting_func_select_bls} plots the weighting function for three baseline lengths: 5, 500, and 2500 m. The left panel plots $\widetilde{\radweight}_{jk}(\eta)$, and the right panel presents the Fourier transformed quantities $\radweight_{jk}(\Delta f)$. To reduce Fourier aliasing effects, we oversample the $\eta$-axis by a factor of 128 compared to the intrinsic sampling given by the data's bandpass extent.

In practice, we represent the weighting function as an $N_\text{freq} \times N_\text{freq}$ matrix $\weightmat_{jk}$ for each baseline $\{j,k\}$ (see Equation \ref{eq:dwcal_matrix}). This allows us to evaluate $\chi^2$ as a matrix multiplication operation. Figure \ref{fig:weighting_matrices} plots two such matrices, for baselines of lengths 5 and 500 m. We generate these matrices for each baseline and pass them to the calibration optimization algorithm.

In \S\ref{s:results} we compare calibration performance using this weighting function to that of sky-based calibration. We implement sky-based calibration by simply setting the weighting matrices equal to the identity matrix for each baseline, i.e.\ $\weightmat_{jk} = \mathbb{1}$.

In order to perform a direct comparison between calibration methods, we desire that the $\chi^2$ do not differ by orders of magnitude. While the overall normalization of the weighting function is arbitrary, vastly different $\chi^2$ magnitudes risk introducing variations in optimization precision. We therefore normalize the weighting functions such that
\begin{equation}
    \sum_{jk} \operatorname{Tr}\big( \weightmat_{jk} \big) = N_\text{bls} N_\text{freq},
\end{equation}
where $N_\text{bls}$ is the number of baselines and $N_\text{freq}$ is the number of frequencies; $\operatorname{Tr}$ denotes the trace.

\subsection{Calibration Optimization}
\label{s:optimization}

We choose to use an off-the-shelf Python-based optimizer for the DWCal explorations presented in this paper. We use \textsc{scipy}'s \texttt{optimize.minimize} function, with the ``method'' option set to ``Newton-CG.'' As this optimizer fits real-valued variables only, we represent the complex gains with their real and imaginary parts.

The inputs to the \texttt{optimize.minimize} function are the data, model visibilities, an initial guess for the gains, and a function that evaluates $\chi^2$. The ``Newton-CG'' minimization method also accepts functions that explicitly calculate Jacobian and Hessian. We analytically calculate those quantities from Equation \ref{eq:dwcal_regularized} and supply them to the optimizer. See \texttt{https://github.com/rlbyrne/dwcal} for those calculations.

We randomize the initial gains, independently selecting the value for each antenna and frequency channel from a complex circular Gaussian distribution with mean value 1 and standard deviation 0.01. We set $\lambda$, the coefficient of the regularization term (see \S\ref{s:regularization}), equal to 0.1.

\begin{table*}
\centering
\begin{tabular}{ | c | c | c | c | c | c | c | }
\hline
 \textbf{Weighting Matrix} & \textbf{Antennas} & \textbf{Baselines} & \textbf{Frequencies} & \textbf{Time Steps} & \textbf{Runtime (minutes)} & \textbf{Gain Error RMS} \\ 
 \hline \hline
 Identity & 127 & 8001 & 384 & 1 & 282 & 0.002705 \\  
 \hline
 Identity & 127 & 8001 & 384 & 2 & 268 & 0.002575 \\
 \hline
 Identity & 127 & 8001 & 384 & 4 & 329 & 0.002538 \\
 \hline
 Weighted & 127 & 8001 & 384 & 1 & 375 & 0.002025 \\
 \hline
 Weighted & 127 & 8001 & 384 & 2 & 473 & 0.001920 \\
 \hline
 Weighted & 127 & 8001 & 384 & 4 & 453 & 0.001882 \\
 \hline
\end{tabular}
\caption{Comparison of six simulated calibration trials. We calibrate with a weighting matrices equal to the identity matrix (analogous to typical sky-based calibration), denoted ``identity'' in the above table, and with the DWCal weighting matrices described in \S\ref{s:weighting_func} and Figure \ref{fig:weighting_matrices}, denoted ``weighted.'' Each calibration trial fits a gain for each of 127 antennas and 384 frequencies, for a total of 48,768 free calibration parameters. The data and model visibilities correspond to 2-minute simulations, and for various calibration trials we average the 2 minute interval into 1, 2, or 4 time steps. In \S\ref{s:results} we present results from the single time step trials. To demonstrate the DWCal concept, we use a highly analytic---but rather slow---optimization method built from \textsc{scipy}'s \texttt{optimize.minimize} function. We find that each optimization operation takes several hours, with the specific convergence timing depending on details of the $\chi^2$ function profile. Future DWCal optimization algorithms are expected to improve optimization efficiency and substantially reduce the runtimes presented here.}
\label{table:run_comparison}
\end{table*}

The primary limitation of this optimization implementation is its speed. Each calibration trial, corresponding to the full 127 antenna array with 384 frequency channels, took several hours to run. In Table \ref{table:run_comparison} we compare six such runs. While the \textsc{scipy} optimizer is sufficient for demonstrating the DWCal concept, we require a more efficient algorithm for calibrating large data volumes.

\subsection{Power Spectrum Estimation}
\label{s:power_spectrum}

After calculating calibration gains, we apply them to the simulated data using \textsc{pyuvdata}'s calibration utility. We then propagate the calibrated data through the \textsc{fhd}/$\epsilon$\textsc{ppsilon} power spectrum estimation pipeline \citep{Sullivan2012, Jacobs2016, Barry2019a, Sullivan2022} to explore the effect of calibration on cosmological power spectrum measurements. \textsc{fhd} grids the visibilities, and $\epsilon$\textsc{ppsilon}\footnote{\texttt{https://github.com/EoRImaging/eppsilon}} calculates error-propagated power spectra. In the next section we present 2D and 1D power spectra generated with $\epsilon$\textsc{ppsilon}.

\section{Results}
\label{s:results}

\begin{figure}
    \centering
    \includegraphics[width=\columnwidth]{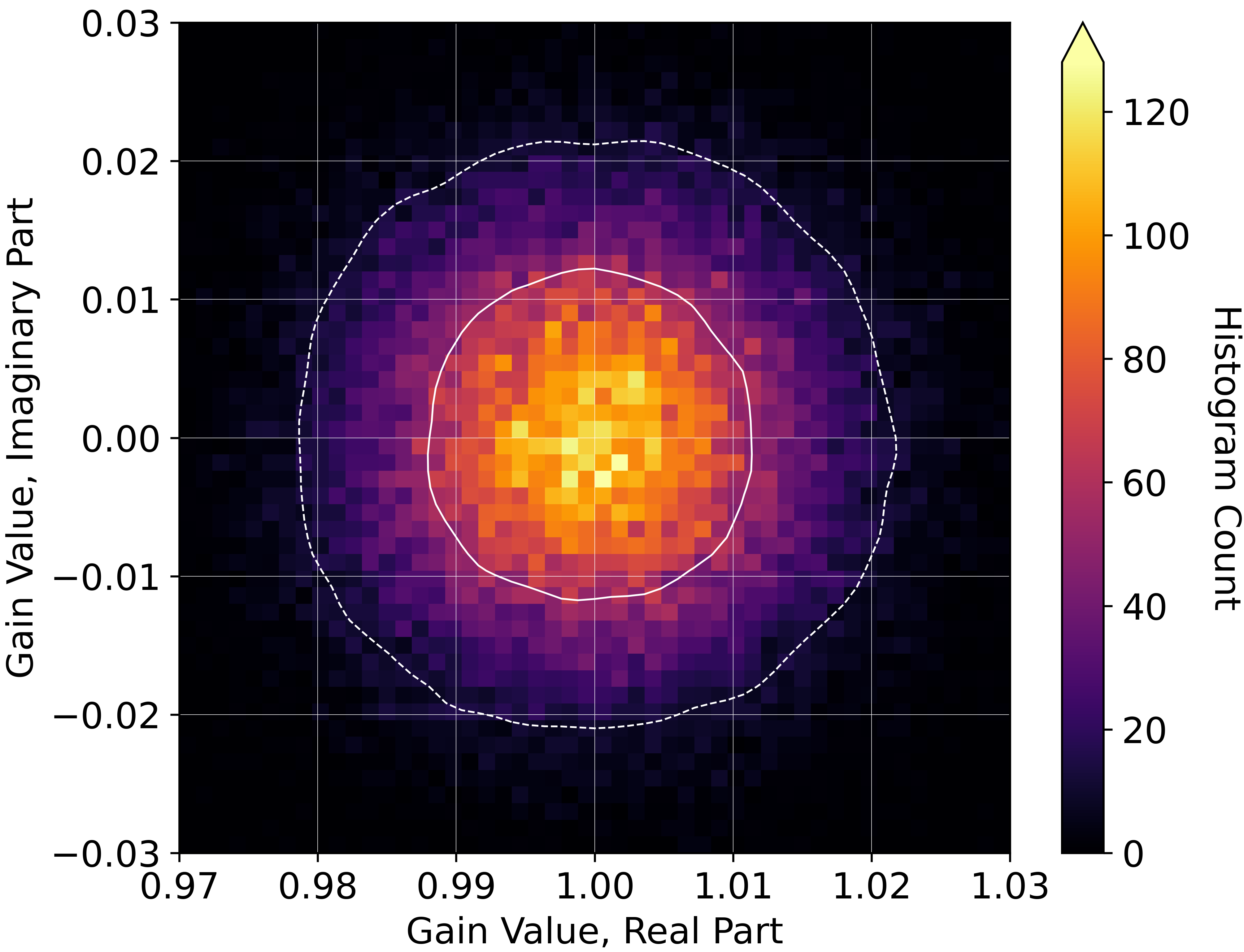}
    \caption{2D histogram of the per-antenna and per-frequency instrumental gains that are applied to the simulated data. The gains are independently drawn from a circular complex Gaussian distribution with mean 1 and standard deviation 0.01 for each antenna and frequency channel. Calibration attempts to recover these gains. We use randomized gains to test the calibration algorithm's robustness against arbitrary complicated instrumental bandpass structure. The solid and dashed white contours enclose an estimated 50\% and 90\% of the distribution, respectively, calculated with a Gaussian kernel density estimator.}
    \label{fig:true_gains_hist}
\end{figure}

We test calibration on simulated data with an instrumental response represented by randomized gains. The gains are selected from a circular complex Gaussian distribution with mean 1 and standard deviation 0.01. We select the values independently for each antenna and frequency channel. To account for the calibration degeneracy in the per-frequency phase (see \S\ref{s:regularization}), we impose a phase rotation at each frequency to ensure that the mean complex phase of the true gains is zero for each frequency channel. Figure \ref{fig:true_gains_hist} plots the distribution of the gains. We divide the simulated visibilities by these gains to derive the data $\data(f)$ used in calibration. Fitting randomized gains in this way ensures that our calibration tests are relevant for instruments with substantial bandpass structure. We show that DWCal performs well for arbitrary instrument responses.

\begin{figure*}
    \centering
    \includegraphics[width=2\columnwidth]{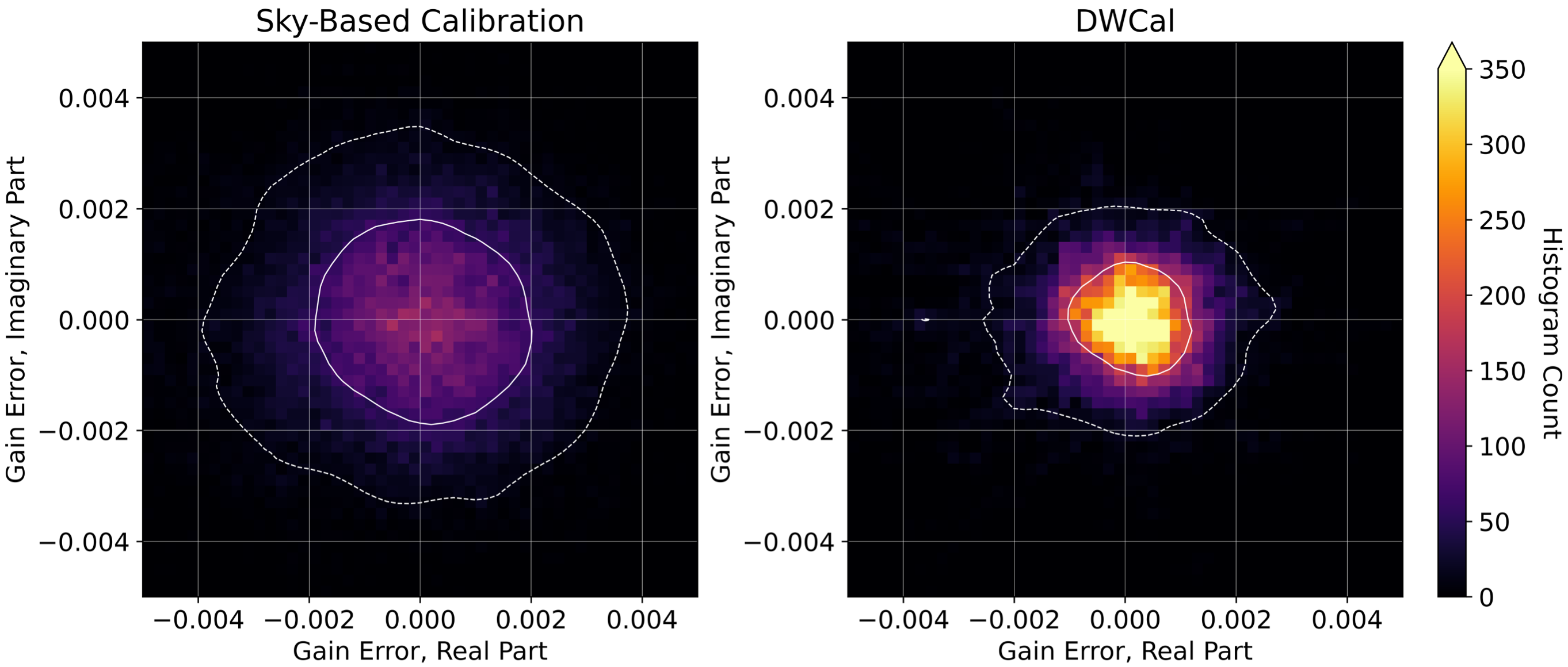}
    \caption{2D histograms of the error on gains fit with sky-based calibration (left) and DWCal (right). DWCal produces more accurate calibration solutions than sky-based calibration. The gains were fit by minimizing Equation \ref{eq:dwcal_regularized} using the optimization procedure described in \S\ref{s:optimization}. For sky-based calibration, the weighting matrices were set equal to the identity matrix, while for DWCal they include off-diagonal elements that downweight modes with substantial sky model error (see \S\ref{s:weighting_func}). The fit gains were compared to the true gains plotted in Figure \ref{fig:true_gains_hist}. Here we plot the discrepancies across all antennas and frequency channels. The solid and dashed contours enclose approximately 50\% and 90\% of the distributions, respectively.}
    \label{fig:gains_error_hist}
\end{figure*}

We calibrate using model visibilities $\modelvals(f)$ simulated from an incomplete sky model (see \S\ref{s:vis_sim}). We calibrate twice, once performing typical sky-based calibration (analogous to setting the weighting matrices equal to the identity matrix) and once implementing DWCal, using the weighting matrices described in \S\ref{s:weighting_func}. Figure \ref{fig:gains_error_hist} plots the distributions of gain error for sky-based calibration (left) and DWCal (right). We find that DWCal produces more accurate calibration solutions than sky-based calibration.

\begin{figure*}
    \centering
    \includegraphics[width=2\columnwidth]{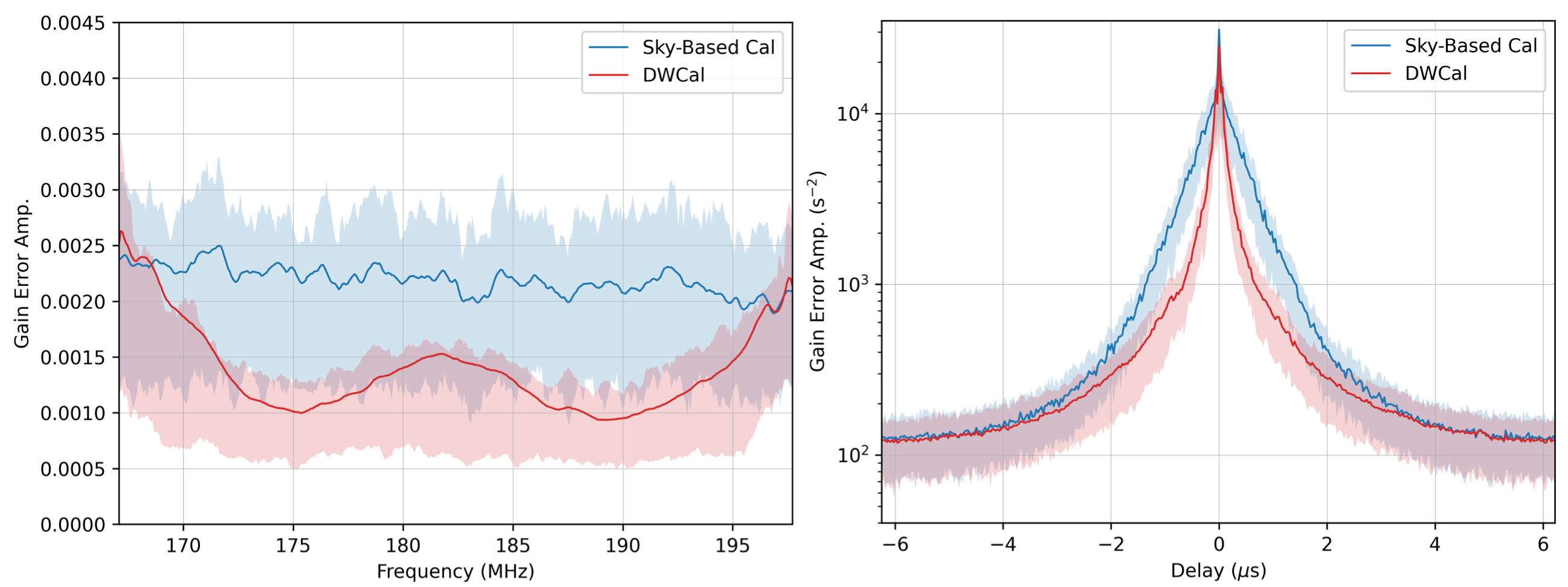}
    \caption{Plot of the gain error as a function of frequency (left) and delay (right). We calculate the gain error by subtracting the true gains (Figure \ref{fig:true_gains_hist}) from the gains fit in calibration. In the left panel, we calculate the amplitude of the gain error and plot it as a function of frequency. In the right panel, we Fourier transform the gain error and plot the amplitude of the result as a function of delay. The solid lines are the antenna-averaged quantities and the shaded regions enclose the central 50\% of values.}
    \label{fig:gains_error}
\end{figure*}

DWCal is fundamentally designed to improve bandpass calibration performance, and it is therefore instructive to explore its impact on the fit gains' frequency dependence. The left panel of Figure \ref{fig:gains_error} plots the amplitude of the fit gain error as a function of frequency. In the right panel we have Fourier transformed the gain error across frequency, and we plot the error amplitude as a function of delay mode. The solid lines indicate the average value across all antennas and the shaded regions enclose the central 50\% of values.

The left panel of Figure \ref{fig:gains_error} indicates that DWCal produces better results than sky-based calibration for most frequencies, with the notable exceptions of the lowest and highest frequency channels. DWCal leverages constraints from neighboring frequency channels to calibrate, and the frequencies on the edges of the band are therefore less constrained. Fortunately, these channels typically contribute minimally to the power spectrum estimate, as analyses generally apply a spectral window function to improve analysis dynamic range (the $\epsilon$\textsc{ppsilon} power spectrum pipeline, for example, applies a Blackman-Harris window function across frequency).

From the right panel of Figure \ref{fig:gains_error} we note that gain error appears predominantly at low delay magnitude. Model visibility error occurs primarily at $\eta=0$. However, the foreground wedge effect extends this error into neighboring delay modes, and sky-based calibration therefore exhibits error in delay modes around $\eta=0$. DWCal does not completely eliminate this effect, but it reduces gain error at these low $|\eta|$.

\begin{figure*}
    \centering
    \includegraphics[width=2\columnwidth]{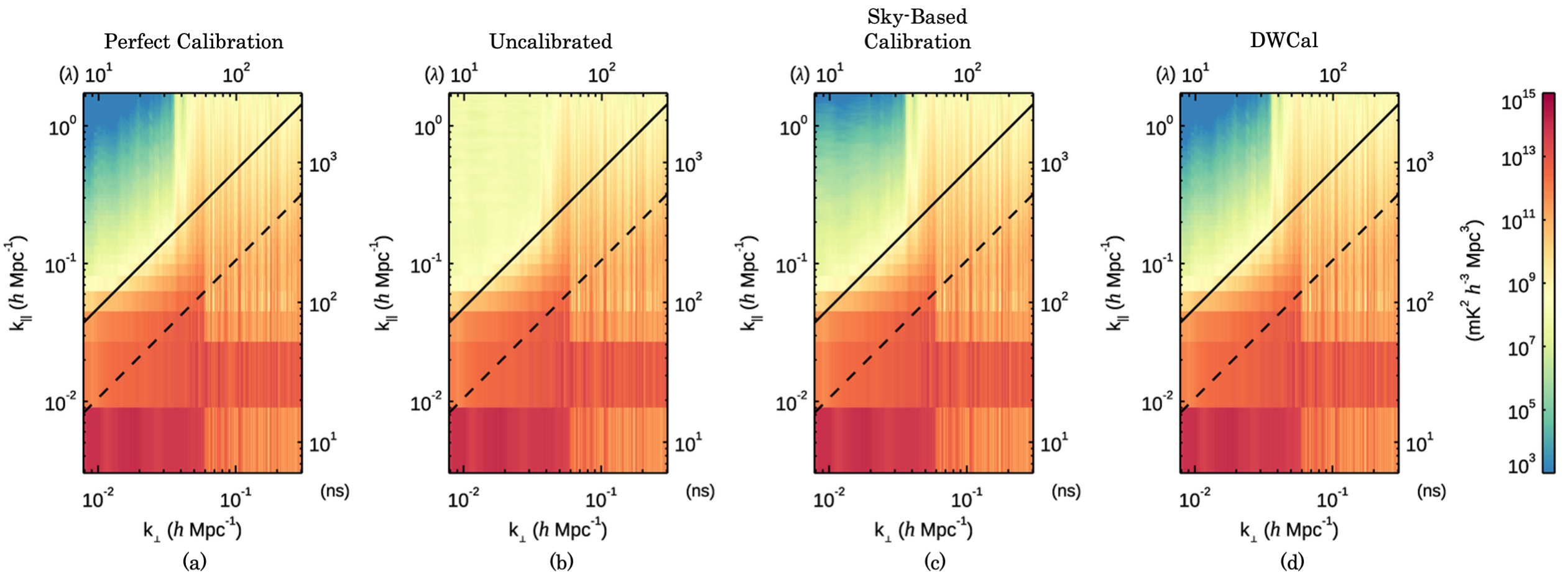}
    \caption{Plots of 2D power spectra generated with the \textsc{fhd}/$\epsilon$\textsc{ppsilon} pipeline. The horizontal axes represent the $k_\perp$ modes perpendicular to the line of sight and the vertical axes represent the $k_\parallel$ modes parallel to the line of sight. Panel (a) corresponds to ``perfect calibration'' and is generated directly from the simulated data. Panel (b), the ``uncalibrated'' power spectrum, has the randomized gains plotted in Figure \ref{fig:true_gains_hist} applied. Panels (c) and (d) have been calibrated with sky-based calibration and DWCal, respectively. In each panel, power is reconstructed predominantly within the foreground wedge. The solid black lines denote the extent of foreground wedge power from sources on the horizon; the dashed black line denotes the extent from sources at the edge of the primary field of view. Beyond the foreground wedge is the EoR window region. The goal of precision calibration is to reduce power in the EoR window. We find that DWCal reconstructs less power in the EoR window than sky-based calibration.}
    \label{fig:2d_ps}
\end{figure*}

To explore the effect of this calibration error on cosmological power spectrum measurements we propagate the visibilities through the \textsc{fhd}/$\epsilon$\textsc{ppsilon} pipeline (see \S\ref{s:power_spectrum}). Figure \ref{fig:2d_ps} depicts the resulting 2D power spectra. Here the horizontal axis denotes modes perpendicular to the line of sight ($k_\perp$) and the vertical axis denotes modes parallel to the line of sight ($k_\parallel$). We present the ``dirty'' power spectra with no foreground subtraction.

The leftmost panel (a) depicts the power spectrum with no calibration error. Here we have used the visibilities directly output from our GLEAM catalog simulation. Power appears predominantly in the foreground wedge region. At high $k_\perp$, poor baseline coverage produces substantial power leakage into high $k_\parallel$. However, at lower $k_\perp$, while we see some leakage into the ``EoR window'' modes (power spectrum modes outside the foreground wedge), the window power is suppressed by many orders of magnitude.

In the second panel (b) of Figure \ref{fig:2d_ps}, we have applied the randomized gains plotted in Figure \ref{fig:true_gains_hist}. This added instrumental response structure couples power into the EoR window, and we see substantially higher power in the window compared to (a).

Panels (c) and (d) of Figure \ref{fig:2d_ps} depict the results of calibrating with sky-based calibration and DWCal, respectively. Here calibration attempts to recover the gains applied in panel (b), reducing power leakage in the EoR window and returning the power spectrum to its state in panel (a). While both methods substantially improve upon the uncalibrated power spectrum in panel (b), DWCal produces less power leakage in the EoR window than sky-based calibration.

\begin{figure}
    \centering
    \includegraphics[width=\columnwidth]{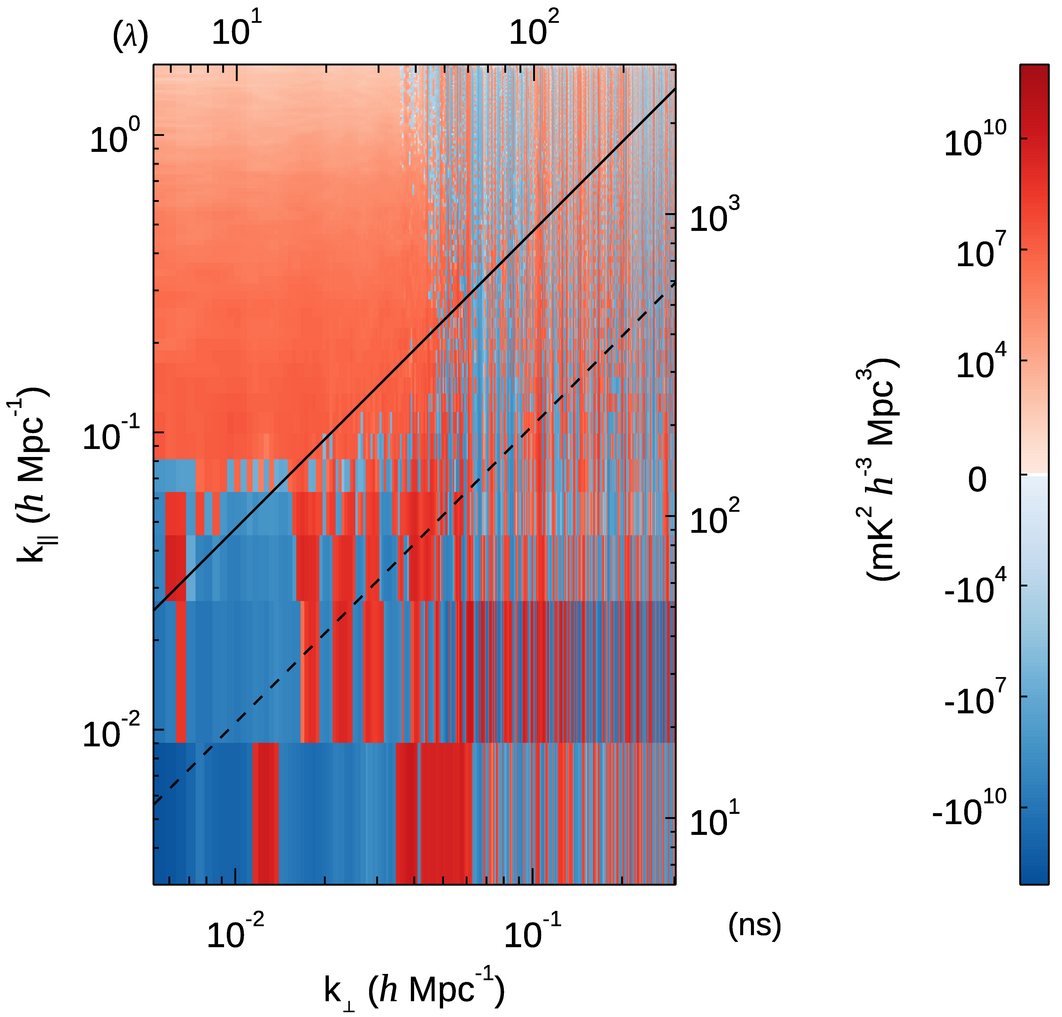}
    \caption{Plot of the 2D power spectrum difference between sky-based calibration and DWCal. Here we have subtracted Figure \ref{fig:2d_ps}(d) from Figure \ref{fig:2d_ps}(c). Modes with positive (red) values have more power when calibrated with sky-based calibration, and modes with negative (blue) values have more power when calibrated with DWCal. We see that DWCal reduces power in nearly all EoR window modes.}
    \label{fig:2d_ps_diff}
\end{figure}

To further highlight the differences between sky-based calibration and DWCal, Figure \ref{fig:2d_ps_diff} plots the difference between Figures \ref{fig:2d_ps}(c) and \ref{fig:2d_ps}(d). Here positive (red) values indicate higher power when calibrated with sky-based calibration, and negative (blue) values indicate higher power when calibrated with DWCal. We see that, throughout the EoR window, sky-based calibration reconstructs more power than DWCal.

\begin{figure}
    \centering
    \includegraphics[width=\columnwidth]{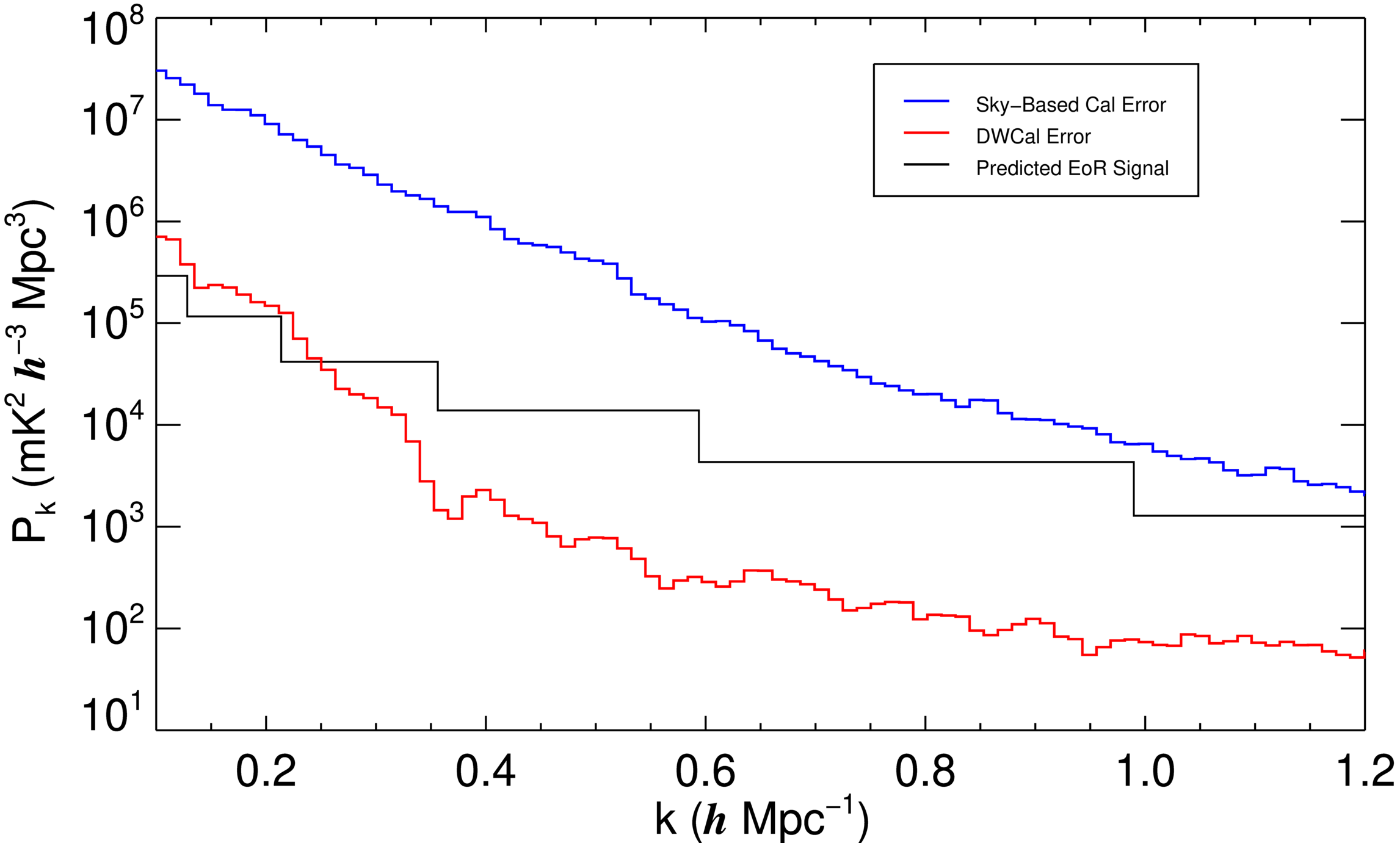}
    \caption{1D power spectra of calibration error from sky-based calibration (blue) and DWCal (red). The black line indicates the predicted Epoch of Reionization (EoR) signal \citep{Furlanetto2006}. The 1D power spectrum is calculated by averaging the EoR window modes of the 2D power spectrum in annuli. Here we use only modes beyond the horizon line. We have subtracted the ``perfect calibration'' power spectrum from Figure \ref{fig:2d_ps}(a) to depict power from calibration error only. We find that DWCal produces a more sensitive power spectrum measurement than sky-based calibration.}
    \label{fig:1d_ps}
\end{figure}

Finally, in Figure \ref{fig:1d_ps} we average the EoR window modes in annuli to compare 1D power spectra produced with sky-based calibration and DWCal. Here we have subtracted the ``perfect calibration'' power spectrum plotted in Figure \ref{fig:2d_ps}(a) to highlight the effect of calibration error only. The power spectrum of sky-based calibration error exceeds that of DWCal by about 2 orders of magnitude. The black line represents the magnitude of the predicted EoR signal \citep{Furlanetto2006}. Sky-based calibration error exceeds the predicted EoR signal on all modes. For the level of sky model error explored here, where we have omitted 10\% of the sky signal from the calibration sky model, sky-based calibration does not enable a detection of the cosmological signal, even in the absence of other systematic error and thermal noise.

Even when calibrating to the same sky model and making no assumptions about the instrumental response, DWCal performs substantially better than sky-based calibration. By incorporating information about the foreground wedge in the form of baseline-dependent weighting matrices, DWCal reduces power leakage into the EoR window and mitigates the impact of calibration error on power spectrum measurements.




\section{Discussion}


DWCal, or delay-weighted calibration, is a new tool for precision bandpass calibration that is resilient to sky model error. Broadly speaking, it is a calibration framework that allows for baseline- and delay-dependent weighting in calibration. This is valuable for 21 cm cosmology experiments, for which sky model error affects a compact region of the 2D power spectrum. DWCal is highly adaptable to a variety of calibration approaches. In this paper we present DWCal as an extension to simple sky-based calibration; however more advanced calibration techniques such as redundant calibration or unified calibration could likewise benefit from the DWCal technique.

DWCal has several distinct benefits over other leading precision calibration techniques. Namely, it makes no assumptions about the antenna gains' bandpass structure, uniformity, or stability. This enables DWCal to fit arbitrary bandpass features and accurately calibrate time-dependent, heterogeneous arrays. DWCal fits the same number of calibration parameters as sky-based calibration and introduces no further calibration degeneracies. Like sky-based calibration, it is degenerate only in the overall phase of the gains at each frequency, and in \S\ref{s:regularization} we introduce an innovative solution for resolving that degeneracy.

The weighting function implemented in DWCal could take any form. Here we present results using the weighting function plotted in Figure \ref{fig:weighting_function}, which fits a simple foreground wedge feature and is designed to approximate the model visibility error calculated in simulation and plotted in Figure \ref{fig:model_error}. We expect DWCal will perform best with weighting functions that most accurately describe the underlying distribution of model visibility error. Future work could explore alternative weighting functions. Model visibility error is highly dependent on the array configuration, antenna beam, and sky models used, so the optimal DWCal weighting function will be unique to each experiment.

This paper demonstrates that DWCal improves calibration's resilience to sky model error. From Figure \ref{fig:1d_ps}, we find that our simulated power spectrum measurements are improved by approximately 2 orders of magnitude on all modes when using DWCal compared to simple sky-based calibration. This moves us from a regime in which calibration error precludes a detection of the 21 cm signal into one in which such a detection would be possible, provided other sources of systematic error are properly controlled. Given that calibration error is one of the dominant barriers to the success of 21 cm experiments, DWCal provides an important tool for enabling these cosmological measurements.

\section*{Acknowledgements}

This work was inspired by conversations with Miguel Morales and Bryna Hazelton. Thank you to Ian Sullivan, Bryna Hazelton, and Nichole Barry for support with the \textsc{fhd}/$\epsilon$\textsc{ppsilon} software pipeline and to Yuping Huang for support with computing resources. Michael Wilensky, Bryna Hazelton, and the anonymous reviewer provided valuable input during revision of this paper.


\bibliography{sample631}{}

\begin{thebibliography}{}
\expandafter\ifx\csname natexlab\endcsname\relax\def\natexlab#1{#1}\fi
\providecommand{\url}[1]{\href{#1}{#1}}
\providecommand{\dodoi}[1]{doi:~\href{http://doi.org/#1}{\nolinkurl{#1}}}
\providecommand{\doeprint}[1]{\href{http://ascl.net/#1}{\nolinkurl{http://ascl.net/#1}}}
\providecommand{\doarXiv}[1]{\href{https://arxiv.org/abs/#1}{\nolinkurl{https://arxiv.org/abs/#1}}}

\bibitem[{Barry {et~al.}(2019{\natexlab{a}})Barry, Beardsley, Byrne, Hazelton,
  Morales, Pober, \& Sullivan}]{Barry2019a}
Barry, N., Beardsley, A.~P., Byrne, R., {et~al.} 2019{\natexlab{a}},
  Publications of the Astronomical Society of Australia, 36, E026,
  \dodoi{10.1017/pasa.2019.21}

\bibitem[{Barry {et~al.}(2016)Barry, Hazelton, Sullivan, Morales, \&
  Pober}]{Barry2016}
Barry, N., Hazelton, B., Sullivan, I., Morales, M.~F., \& Pober, J.~C. 2016,
  Monthly Notices of the Royal Astronomical Society, 461, 3135,
  \dodoi{10.1093/mnras/stw1380}

\bibitem[{Barry {et~al.}(2019{\natexlab{b}})Barry, Wilensky, Trott, Pindor,
  Beardsley, Hazelton, Sullivan, Morales, Pober, Line, Greig, Byrne, Lanman,
  Li, Jordan, Joseph, McKinley, Rahimi, Yoshiura, Bowman, Gaensler, Hewitt,
  Jacobs, Mitchell, {Udaya Shankar}, Sethi, Subrahmanyan, Tingay, Webster, \&
  Wyithe}]{Barry2019b}
Barry, N., Wilensky, M., Trott, C.~M., {et~al.} 2019{\natexlab{b}}, The
  Astrophysical Journal, 884, 1, \dodoi{10.3847/1538-4357/ab40a8}

\bibitem[{Byrne(2022)}]{DWCal2022}
Byrne, R. 2022, {DWCal},  Zenodo, \dodoi{10.5281/zenodo.7402471}

\bibitem[{Byrne {et~al.}(2022)Byrne, Morales, Hazelton, Sullivan, \&
  Barry}]{Byrne2022a}
Byrne, R., Morales, M.~F., Hazelton, B., Sullivan, I., \& Barry, N. 2022,
  Publications of the Astronomical Society of Australia, 39, e023,
  \dodoi{10.1017/pasa.2022.21}

\bibitem[{Byrne {et~al.}(2021{\natexlab{a}})Byrne, Morales, Hazelton, Sullivan,
  Barry, Lynch, Line, \& Jacobs}]{Byrne2021c}
Byrne, R., Morales, M.~F., Hazelton, B., {et~al.} 2021{\natexlab{a}}, Monthly
  Notices of the Royal Astronomical Society, 510, 2011,
  \dodoi{10.1093/mnras/stab3276}

\bibitem[{Byrne {et~al.}(2021{\natexlab{b}})Byrne, Morales, Hazelton, \&
  Wilensky}]{Byrne2021b}
Byrne, R., Morales, M.~F., Hazelton, B.~J., \& Wilensky, M. 2021{\natexlab{b}},
  Monthly Notices of the Royal Astronomical Society, 503, 2457,
  \dodoi{10.1093/mnras/stab647}

\bibitem[{Byrne {et~al.}(2019)Byrne, Morales, Hazelton, Li, Barry, Beardsley,
  Joseph, Pober, Sullivan, \& Trott}]{Byrne2019}
Byrne, R., Morales, M.~F., Hazelton, B., {et~al.} 2019, The Astrophysical
  Journal, 875, 70, \dodoi{10.3847/1538-4357/ab107d}

\bibitem[{Carroll {et~al.}(2016)Carroll, Line, Morales, Barry, Beardsley,
  Hazelton, Jacobs, Pober, Sullivan, Webster, Bernardi, Bowman, Briggs,
  Cappallo, Corey, de~Oliveira-Costa, Dillon, Emrich, Ewall-Wice, Feng,
  Gaensler, Goeke, Greenhill, Hewitt, Hurley-Walker, Johnston-Hollitt, Kaplan,
  Kasper, Kim, Kratzenberg, Lenc, Loeb, Lonsdale, Lynch, McKinley, McWhirter,
  Mitchell, Morgan, Neben, Oberoi, Offringa, Ord, Paul, Pindor, Prabu,
  Procopio, Riding, Rogers, Roshi, Shankar, Sethi, Srivani, Subrahmanyan,
  Tegmark, Thyagarajan, Tingay, Trott, Waterson, Wayth, Whitney, Williams,
  Williams, Wu, \& Wyithe}]{Carroll2016}
Carroll, P.~A., Line, J., Morales, M.~F., {et~al.} 2016, Monthly Notices of the
  Royal Astronomical Society, 461, 4151, \dodoi{10.1093/mnras/stw1599}

\bibitem[{de~Gasperin {et~al.}(2020)de~Gasperin, Vink, McKean, Asgekar, Avruch,
  Bentum, Blaauw, Bonafede, Broderick, Br{\"{u}}ggen, Breitling, Brouw,
  Butcher, Ciardi, Cuciti, de~Vos, Duscha, Eisl{\"{o}}ffel, Engels, Fallows,
  Franzen, Garrett, Gunst, H{\"{o}}randel, Heald, Hoeft, Iacobelli, Koopmans,
  Krankowski, Maat, Mann, Mevius, Miley, Morganti, Nelles, Norden, Offringa,
  Orr{\'{u}}, Paas, Pandey, Pandey-Pommier, Pekal, Pizzo, Reich, Rowlinson,
  Rottgering, Schwarz, Shulevski, Smirnov, Sobey, Soida, Steinmetz, Tagger,
  Toribio, van Ardenne, van~der Horst, van Haarlem, van Weeren, Vocks,
  Wucknitz, Zarka, \& Zucca}]{deGasperin2020}
de~Gasperin, F., Vink, J., McKean, J.~P., {et~al.} 2020, Astronomy \&
  Astrophysics, 635, A150, \dodoi{10.1051/0004-6361/201936844}

\bibitem[{Dillon {et~al.}(2015)Dillon, Tegmark, Liu, Ewall-Wice, Hewitt,
  Morales, Neben, Parsons, \& Zheng}]{Dillon2015}
Dillon, J.~S., Tegmark, M., Liu, A., {et~al.} 2015, Physical Review D, 91,
  \dodoi{10.1103/physrevd.91.023002}

\bibitem[{Dillon {et~al.}(2018)Dillon, Kohn, Parsons, Aguirre, Ali, Bernardi,
  Kern, Li, Liu, Nunhokee, \& Pober}]{Dillon2018}
Dillon, J.~S., Kohn, S.~A., Parsons, A.~R., {et~al.} 2018, Monthly Notices of
  the Royal Astronomical Society, 477, 5670, \dodoi{10.1093/mnras/sty1060}

\bibitem[{Dillon {et~al.}(2020)Dillon, Lee, Ali, Parsons, Orosz, Nunhokee, {La
  Plante}, Beardsley, Kern, Abdurashidova, Aguirre, Alexander, Balfour,
  Bernardi, Billings, Bowman, Bradley, Bull, Burba, Carey, Carilli, Cheng,
  DeBoer, Dexter, {de Lera Acedo}, Ely, Ewall-Wice, Fagnoni, Fritz, Furlanetto,
  Gale-Sides, Glendenning, Gorthi, Greig, Grobbelaar, Halday, Hazelton, Hewitt,
  Hickish, Jacobs, Julius, Kerrigan, Kittiwisit, Kohn, Kolopanis, Lanman,
  Lekalake, Lewis, Liu, Ma, MacMahon, Malan, Malgas, Maree, Martinot,
  Matsetela, Mesinger, Molewa, Morales, Mosiane, Murray, Neben, Nikolic,
  Pascua, Patra, Pieterse, Pober, Razavi-Ghods, Ringuette, Robnett, Rosie,
  Santos, Sims, Smith, Syce, Tegmark, Thyagarajan, Williams, \&
  Zheng}]{Dillon2020}
Dillon, J.~S., Lee, M., Ali, Z.~S., {et~al.} 2020, Monthly Notices of the Royal
  Astronomical Society, 499, 5840, \dodoi{10.1093/mnras/staa3001}

\bibitem[{Eastwood {et~al.}(2018)Eastwood, Anderson, Monroe, Hallinan,
  Barsdell, Bourke, Clark, Ellingson, Dowell, Garsden, Greenhill, Hartman,
  Kocz, Lazio, Price, Schinzel, Taylor, Vedantham, Wang, \&
  Woody}]{Eastwood2018}
Eastwood, M.~W., Anderson, M.~M., Monroe, R.~M., {et~al.} 2018, The
  Astronomical Journal, 156, 32, \dodoi{10.3847/1538-3881/aac721}

\bibitem[{Eastwood {et~al.}(2019)Eastwood, Anderson, Monroe, Hallinan, Catha,
  Dowell, Garsden, Greenhill, Hicks, Kocz, \& al.}]{Eastwood2019}
---. 2019, The Astronomical Journal, 158, 84, \dodoi{10.3847/1538-3881/ab2629}

\bibitem[{Ewall-Wice {et~al.}(2022)Ewall-Wice, Dillon, Gehlot, Parsons, Cox, \&
  Jacobs}]{ewall-wice2022}
Ewall-Wice, A., Dillon, J.~S., Gehlot, B., {et~al.} 2022, The Astrophysical
  Journal, 938, 151, \dodoi{10.3847/1538-4357/ac87b3}

\bibitem[{Ewall-Wice {et~al.}(2016)Ewall-Wice, Dillon, Liu, \&
  Hewitt}]{Ewall-Wice2017}
Ewall-Wice, A., Dillon, J.~S., Liu, A., \& Hewitt, J. 2016, Monthly Notices of
  the Royal Astronomical Society, 470, 1849, \dodoi{10.1093/mnras/stx1221}

\bibitem[{Furlanetto {et~al.}(2006)Furlanetto, {Peng Oh}, \&
  Briggs}]{Furlanetto2006}
Furlanetto, S.~R., {Peng Oh}, S., \& Briggs, F.~H. 2006, Physics Reports, 433,
  181, \dodoi{10.1016/j.physrep.2006.08.002}

\bibitem[{Gehlot {et~al.}(2018)Gehlot, Koopmans, de~Bruyn, Zaroubi, Brentjens,
  Asad, Hatef, Jeli{\'{c}}, Mevius, Offringa, Pandey, \&
  Yatawatta}]{Gehlot2018}
Gehlot, B.~K., Koopmans, L.~V., de~Bruyn, A.~G., {et~al.} 2018, Monthly Notices
  of the Royal Astronomical Society, 478, 1484, \dodoi{10.1093/mnras/sty1095}

\bibitem[{Grobler {et~al.}(2018)Grobler, Bernardi, Kenyon, Parsons, \&
  Smirnov}]{Grobler2018}
Grobler, T.~L., Bernardi, G., Kenyon, J.~S., Parsons, A.~R., \& Smirnov, O.~M.
  2018, Monthly Notices of the Royal Astronomical Society, 476, 2410,
  \dodoi{10.1093/mnras/sty357}

\bibitem[{Hamaker {et~al.}(1996)Hamaker, Bregman, \& Sault}]{Hamaker1996a}
Hamaker, J.~P., Bregman, J.~D., \& Sault, R.~J. 1996, Astronomy and
  Astrophysics Supplement Series, 117, 137, \dodoi{10.1051/aas:1996146}

\bibitem[{Hazelton {et~al.}(2017)Hazelton, Jacobs, Pober, \&
  Beardsley}]{Hazelton2017}
Hazelton, B.~J., Jacobs, D.~C., Pober, J.~C., \& Beardsley, A.~P. 2017, The
  Journal of Open Source Software, 2, 140, \dodoi{10.21105/joss.00140}

\bibitem[{Hazelton {et~al.}(2013)Hazelton, Morales, \& Sullivan}]{Hazelton2013}
Hazelton, B.~J., Morales, M.~F., \& Sullivan, I.~S. 2013, The Astrophysical
  Journal, 770, 156, \dodoi{10.1088/0004-637X/770/2/156}

\bibitem[{Hurley-Walker {et~al.}(2017)Hurley-Walker, Callingham, Hancock,
  Franzen, Hindson, Kapi{\'{n}}ska, Morgan, Offringa, Wayth, Wu, Zheng, Murphy,
  Bell, Dwarakanath, For, Gaensler, Johnston-Hollitt, Lenc, Procopio,
  Staveley-Smith, Ekers, Bowman, Briggs, Cappallo, Deshpande, Greenhill,
  Hazelton, Kaplan, Lonsdale, McWhirter, Mitchell, Morales, Morgan, Oberoi,
  Ord, Prabu, {Udaya Shankar}, Srivani, Subrahmanyan, Tingay, Webster,
  Williams, \& Williams}]{Hurley-Walker2017}
Hurley-Walker, N., Callingham, J.~R., Hancock, P.~J., {et~al.} 2017, Monthly
  Notices of the Royal Astronomical Society, 464, 1146,
  \dodoi{10.1093/mnras/stw2337}

\bibitem[{Jacobs {et~al.}(2016)Jacobs, Hazelton, Trott, Dillon, Pindor,
  Sullivan, Pober, Barry, Beardsley, Bernardi, Bowman, Briggs, Cappallo,
  Carroll, Corey, de~Oliveira-Costa, Emrich, Ewall-Wice, Feng, Gaensler, Goeke,
  Greenhill, Hewitt, Hurley-Walker, Johnston-Hollitt, Kaplan, Kasper, Kim,
  Kratzenberg, Lenc, Line, Loeb, Lonsdale, Lynch, McKinley, McWhirter,
  Mitchell, Morales, Morgan, Neben, Thyagarajan, Oberoi, Offringa, Ord, Paul,
  Prabu, Procopio, Riding, Rogers, Roshi, Shankar, Sethi, Srivani,
  Subrahmanyan, Tegmark, Tingay, Waterson, Wayth, Webster, Whitney, Williams,
  Williams, Wu, \& Wyithe}]{Jacobs2016}
Jacobs, D.~C., Hazelton, B.~J., Trott, C.~M., {et~al.} 2016, The Astrophysical
  Journal, 825, 114, \dodoi{10.3847/0004-637X/825/2/114}

\bibitem[{Kazemi \& Yatawatta(2013)}]{Kazemi2013}
Kazemi, S., \& Yatawatta, S. 2013, Monthly Notices of the Royal Astronomical
  Society, 435, 597, \dodoi{10.1093/mnras/stt1347}

\bibitem[{Kazemi {et~al.}(2011)Kazemi, Yatawatta, Zaroubi, Lampropoulos,
  de~Bruyn, Koopmans, \& Noordam}]{Kazemi2011}
Kazemi, S., Yatawatta, S., Zaroubi, S., {et~al.} 2011, Monthly Notices of the
  Royal Astronomical Society, 414, 1656,
  \dodoi{10.1111/j.1365-2966.2011.18506.x}

\bibitem[{Kern {et~al.}(2020)Kern, Dillon, Parsons, Carilli, Bernardi,
  Abdurashidova, Aguirre, Alexander, Ali, Balfour, Beardsley, Billings, Bowman,
  Bradley, Bull, Burba, Carey, Cheng, DeBoer, Dexter, {de Lera Acedo}, Ely,
  Ewall-Wice, Fagnoni, Fritz, Furlanetto, Gale-Sides, Glendenning, Gorthi,
  Greig, Grobbelaar, Halday, Hazelton, Hewitt, Hickish, Jacobs, Julius,
  Kerrigan, Kittiwisit, Kohn, Kolopanis, Lanman, {La Plante}, Lekalake, Liu,
  MacMahon, Malan, Malgas, Maree, Martinot, Matsetela, Mesinger, Molewa,
  Morales, Mosiane, Murray, Neben, Nikolic, Nunhokee, Patra, Pieterse, Pober,
  Razavi-Ghods, Ringuette, Robnett, Rosie, Sims, Smith, Syce, Thyagarajan,
  Williams, \& Zheng}]{Kern2020}
Kern, N.~S., Dillon, J.~S., Parsons, A.~R., {et~al.} 2020, The Astrophysical
  Journal, 890, 122, \dodoi{10.3847/1538-4357/ab67bc}

\bibitem[{Li {et~al.}(2018)Li, Pober, Hazelton, Barry, Morales, Sullivan,
  Parsons, Ali, Dillon, Beardsley, Bowman, Briggs, Byrne, Carroll, Crosse,
  Emrich, Ewall-Wice, Feng, Franzen, Hewitt, Horsley, Jacobs, Johnston-Hollitt,
  Jordan, Joseph, Kaplan, Kenney, Kim, Kittiwisit, Lanman, Line, McKinley,
  Mitchell, Murray, Neben, Offringa, Pallot, Paul, Pindor, Procopio, Rahimi,
  Riding, Sethi, {Udaya Shankar}, Steele, Subrahmanian, Tegmark, Thyagarajan,
  Tingay, Trott, Walker, Wayth, Webster, Williams, Wu, \& Wyithe}]{Li2018}
Li, W., Pober, J.~C., Hazelton, B.~J., {et~al.} 2018, The Astrophysical
  Journal, 863, 170, \dodoi{10.3847/1538-4357/aad3c3}

\bibitem[{Li {et~al.}(2019)Li, Pober, Barry, Hazelton, Morales, Trott, Lanman,
  Wilensky, Sullivan, Beardsley, Booler, Bowman, Byrne, Crosse, Emrich,
  Franzen, Hasegawa, Horsley, Johnston-Hollitt, Jacobs, Jordan, Joseph,
  Kaneuji, Kaplan, Kenney, Kubota, Line, Lynch, McKinley, Mitchell, Murray,
  Pallot, Pindor, Rahimi, Riding, Sleap, Steele, Takahashi, Tingay, Walker,
  Wayth, Webster, Williams, Wu, Wyithe, Yoshiura, \& Zheng}]{Li2019}
Li, W., Pober, J.~C., Barry, N., {et~al.} 2019, The Astrophysical Journal, 887,
  141, \dodoi{10.3847/1538-4357/ab55e4}

\bibitem[{Liu \& Shaw(2020)}]{Liu2020}
Liu, A., \& Shaw, J.~R. 2020, Publications of the Astronomical Society of the
  Pacific, 132, 062001, \dodoi{10.1088/1538-3873/ab5bfd}

\bibitem[{Liu {et~al.}(2010)Liu, Tegmark, Morrison, Lutomirski, \&
  Zaldarriaga}]{Liu2010}
Liu, A., Tegmark, M., Morrison, S., Lutomirski, A., \& Zaldarriaga, M. 2010,
  Monthly Notices of the Royal Astronomical Society, 408, 1029,
  \dodoi{10.1111/j.1365-2966.2010.17174.x}

\bibitem[{Mertens {et~al.}(2020)Mertens, Mevius, Koopmans, Offringa, Mellema,
  Zaroubi, Brentjens, Gan, Gehlot, Pandey, Sardarabadi, Vedantham, Yatawatta,
  Asad, Ciardi, Chapman, Gazagnes, Ghara, Ghosh, Giri, Iliev, Jeli{\'{c}},
  Kooistra, Mondal, Schaye, \& Silva}]{Mertens2020}
Mertens, F.~G., Mevius, M., Koopmans, L. V.~E., {et~al.} 2020, Monthly Notices
  of the Royal Astronomical Society, 493, 1662, \dodoi{10.1093/mnras/staa327}

\bibitem[{Mitchell {et~al.}(2008)Mitchell, Greenhill, Wayth, Sault, Lonsdale,
  Cappallo, Morales, \& Ord}]{Mitchell2008}
Mitchell, D., Greenhill, L., Wayth, R., {et~al.} 2008, IEEE Journal of Selected
  Topics in Signal Processing, 2, 707, \dodoi{10.1109/JSTSP.2008.2005327}

\bibitem[{Morales {et~al.}(2019)Morales, Beardsley, Pober, Barry, Hazelton,
  Jacobs, \& Sullivan}]{Morales2019}
Morales, M.~F., Beardsley, A., Pober, J., {et~al.} 2019, Monthly Notices of the
  Royal Astronomical Society, 483, 2207, \dodoi{10.1093/mnras/sty2844}

\bibitem[{Morales {et~al.}(2012)Morales, Hazelton, Sullivan, \&
  Beardsley}]{Morales2012}
Morales, M.~F., Hazelton, B., Sullivan, I., \& Beardsley, A. 2012,
  Astrophysical Journal, 752, 137, \dodoi{10.1088/0004-637X/752/2/137}

\bibitem[{Morales \& Wyithe(2010)}]{Morales2010}
Morales, M.~F., \& Wyithe, J. S.~B. 2010, Annual Review of Astronomy and
  Astrophysics, 48, 127, \dodoi{10.1146/annurev-astro-081309-130936}

\bibitem[{Offringa {et~al.}(2015)Offringa, Wayth, Hurley-Walker, Kaplan, Barry,
  Beardsley, Bell, Bernardi, Bowman, Briggs, Callingham, Cappallo, Carroll,
  Deshpande, Dillon, Dwarakanath, Ewall-Wice, Feng, For, Gaensler, Greenhill,
  Hancock, Hazelton, Hewitt, Hindson, Jacobs, Johnston-Hollitt, Kapi{\'{n}}ska,
  Kim, Kittiwisit, Lenc, Line, Loeb, Lonsdale, McKinley, McWhirter, Mitchell,
  Morales, Morgan, Morgan, Neben, Oberoi, Ord, Paul, Pindor, Pober, Prabu,
  Procopio, Riding, {Udaya Shankar}, Sethi, Srivani, Staveley-Smith,
  Subrahmanyan, Sullivan, Tegmark, Thyagarajan, Tingay, Trott, Webster,
  Williams, Williams, Wu, Wyithe, \& Zheng}]{Offringa2015}
Offringa, A., Wayth, R., Hurley-Walker, N., {et~al.} 2015, Publications of the
  Astronomical Society of Australia, 32, E008, \dodoi{10.1017/pasa.2015.7}

\bibitem[{Parsons \& Backer(2009)}]{Parsons2009}
Parsons, A.~R., \& Backer, D.~C. 2009, The Astronomical Journal, 138, 219,
  \dodoi{10.1088/0004-6256/138/1/219}

\bibitem[{Parsons {et~al.}(2012)Parsons, Pober, Aguirre, Carilli, Jacobs, \&
  Moore}]{Parsons2012}
Parsons, A.~R., Pober, J.~C., Aguirre, J.~E., {et~al.} 2012, Astrophysical
  Journal, 756, 165, \dodoi{10.1088/0004-637X/756/2/165}

\bibitem[{Patil {et~al.}(2017)Patil, Yatawatta, Koopmans, de~Bruyn, Brentjens,
  Zaroubi, Asad, Hatef, Jeli{\'{c}}, Mevius, Offringa, Pandey, Vedantham,
  Abdalla, Brouw, Chapman, Ciardi, Gehlot, Ghosh, Harker, Iliev, Kakiichi,
  Majumdar, Mellema, Silva, Schaye, Vrbanec, \& Wijnholds}]{Patil2017}
Patil, A.~H., Yatawatta, S., Koopmans, L. V.~E., {et~al.} 2017, The
  Astrophysical Journal, 838, 65, \dodoi{10.3847/1538-4357/aa63e7}

\bibitem[{Pober {et~al.}(2013)Pober, Parsons, Aguirre, Ali, Bradley, Carilli,
  DeBoer, Dexter, Gugliucci, Jacobs, Klima, MacMahon, Manley, Moore, Stefan, \&
  Walbrugh}]{Pober2013}
Pober, J.~C., Parsons, A.~R., Aguirre, J.~E., {et~al.} 2013, The Astrophysical
  Journal, 768, L36, \dodoi{10.1088/2041-8205/768/2/L36}

\bibitem[{Pritchard \& Loeb(2008)}]{Pritchard2008}
Pritchard, J.~R., \& Loeb, A. 2008, Physical Review D, 78, 103511,
  \dodoi{10.1103/PhysRevD.78.103511}

\bibitem[{Salvini \& Wijnholds(2014)}]{Salvini2014}
Salvini, S., \& Wijnholds, S.~J. 2014, in 2014 31th URSI General Assembly and
  Scientific Symposium, URSI GASS 2014, \dodoi{10.1109/URSIGASS.2014.6930038}

\bibitem[{Sault {et~al.}(1996)Sault, Hamaker, \& Bregman}]{Sault1996}
Sault, R.~J., Hamaker, J.~P., \& Bregman, J.~D. 1996, Astronomy and
  Astrophysics Supplement Series, 117, 149, \dodoi{10.1051/aas:1996100}

\bibitem[{Sievers(2017)}]{Sievers2017}
Sievers, J.~L. 2017, arXiv e-prints.
\newblock \doarXiv{1701.01860}

\bibitem[{Sims {et~al.}(2022{\natexlab{a}})Sims, Pober, \& Sievers}]{Sims2022}
Sims, P.~H., Pober, J.~C., \& Sievers, J.~L. 2022{\natexlab{a}}, Monthly
  Notices of the Royal Astronomical Society, 517, 910,
  \dodoi{10.1093/mnras/stac1861}

\bibitem[{Sims {et~al.}(2022{\natexlab{b}})Sims, Pober, \& Sievers}]{Sims2022b}
---. 2022{\natexlab{b}}, Monthly Notices of the Royal Astronomical Society,
  517, 935, \dodoi{10.1093/mnras/stac1749}

\bibitem[{Sob {et~al.}(2020)Sob, Bester, Smirnov, Kenyon, \& Grobler}]{Sob2020}
Sob, U.~M., Bester, H.~L., Smirnov, O.~M., Kenyon, J.~S., \& Grobler, T.~L.
  2020, Monthly Notices of the Royal Astronomical Society, 491, 1026,
  \dodoi{10.1093/mnras/stz3037}

\bibitem[{Sullivan {et~al.}(2022)Sullivan, Barry, Byrne, Morales, Hazelton,
  Beardsley, \& Lanman}]{Sullivan2022}
Sullivan, I., Barry, N., Byrne, R.~L., {et~al.} 2022, Astrophysics Source Code
  Library.
\newblock \url{https://ui.adsabs.harvard.edu/abs/2022ascl.soft05014S}

\bibitem[{Sullivan {et~al.}(2012)Sullivan, Morales, Hazelton, Arcus, Barnes,
  Bernardi, Briggs, Bowman, Bunton, Cappallo, Corey, Deshpande, Desouza,
  Emrich, Gaensler, Goeke, Greenhill, Herne, Hewitt, Johnston-Hollitt, Kaplan,
  Kasper, Kincaid, Koenig, Kratzenberg, Lonsdale, Lynch, McWhirter, Mitchell,
  Morgan, Oberoi, Ord, Pathikulangara, Prabu, Remillard, Rogers, Roshi, Salah,
  Sault, Shankar, Srivani, Stevens, Subrahmanyan, Tingay, Wayth, Waterson,
  Webster, Whitney, Williams, Williams, \& Wyithe}]{Sullivan2012}
Sullivan, I.~S., Morales, M.~F., Hazelton, B.~J., {et~al.} 2012, Astrophysical
  Journal, 759, 17, \dodoi{10.1088/0004-637X/759/1/17}

\bibitem[{Sutinjo {et~al.}(2015)Sutinjo, O'Sullivan, Lenc, Wayth, Padhi, Hall,
  \& Tingay}]{Sutinjo2015}
Sutinjo, A., O'Sullivan, J., Lenc, E., {et~al.} 2015, Radio Science, 50, 52,
  \dodoi{10.1002/2014RS005517}

\bibitem[{Thyagarajan {et~al.}(2013)Thyagarajan, {Udaya Shankar}, Subrahmanyan,
  Arcus, Bernardi, Bowman, Briggs, Bunton, Cappallo, Corey, Desouza, Emrich,
  Gaensler, Goeke, Greenhill, Hazelton, Herne, Hewitt, Johnston-Hollitt,
  Kaplan, Kasper, Kincaid, Koenig, Kratzenberg, Lonsdale, Lynch, McWhirter,
  Mitchell, Morales, Morgan, Oberoi, Ord, Pathikulangara, Remillard, Rogers,
  Roshi, Salah, Sault, Srivani, Stevens, Thiagaraj, Tingay, Wayth, Waterson,
  Webster, Whitney, Williams, Williams, \& Wyithe}]{Thyagarajan2013}
Thyagarajan, N., {Udaya Shankar}, N., Subrahmanyan, R., {et~al.} 2013,
  Astrophysical Journal, 776, 6, \dodoi{10.1088/0004-637X/776/1/6}

\bibitem[{Tingay {et~al.}(2013)Tingay, Goeke, Bowman, Emrich, Ord, Mitchell,
  Morales, Booler, Crosse, Wayth, Lonsdale, Tremblay, Pallot, Colegate,
  Wicenec, Kudryavtseva, Arcus, Barnes, Bernardi, Briggs, Burns, Bunton,
  Cappallo, Corey, Deshpande, Desouza, Gaensler, Greenhill, Hall, Hazelton,
  Herne, Hewitt, Johnston-Hollitt, Kaplan, Kasper, Kincaid, Koenig,
  Kratzenberg, Lynch, McKinley, McWhirter, Morgan, Oberoi, Pathikulangara,
  Prabu, Remillard, Rogers, Roshi, Salah, Sault, Udaya-Shankar, Schlagenhaufer,
  Srivani, Stevens, Subrahmanyan, Waterson, Webster, Whitney, Williams,
  Williams, \& Wyithe}]{Tingay2013}
Tingay, S.~J., Goeke, R., Bowman, J.~D., {et~al.} 2013, Publications of the
  Astronomical Society of Australia, 30, E007, \dodoi{10.1017/pasa.2012.007}

\bibitem[{Trott {et~al.}(2012)Trott, Wayth, \& Tingay}]{Trott2012}
Trott, C.~M., Wayth, R.~B., \& Tingay, S.~J. 2012, Astrophysical Journal, 757,
  101, \dodoi{10.1088/0004-637X/757/1/101}

\bibitem[{Trott {et~al.}(2016)Trott, Pindor, Procopio, Wayth, Mitchell,
  McKinley, Tingay, Barry, Beardsley, Bernardi, Bowman, Briggs, Cappallo,
  Carroll, de~Oliveira-Costa, Dillon, Ewall-Wice, Feng, Greenhill, Hazelton,
  Hewitt, Hurley-Walker, Johnston-Hollitt, Jacobs, Kaplan, Kim, Lenc, Line,
  Loeb, Lonsdale, Morales, Morgan, Neben, Thyagarajan, Oberoi, Offringa, Ord,
  Paul, Pober, Prabu, Riding, Shankar, Sethi, Srivani, Subrahmanyan, Sullivan,
  Tegmark, Webster, Williams, Williams, Wu, \& Wyithe}]{Trott2016}
Trott, C.~M., Pindor, B., Procopio, P., {et~al.} 2016, The Astrophysical
  Journal, 818, 139, \dodoi{10.3847/0004-637x/818/2/139}

\bibitem[{Vedantham {et~al.}(2012)Vedantham, {Udaya Shankar}, \&
  Subrahmanyan}]{Vedantham2012}
Vedantham, H., {Udaya Shankar}, N., \& Subrahmanyan, R. 2012, Astrophysical
  Journal, 745, 176, \dodoi{10.1088/0004-637X/745/2/176}

\bibitem[{Wieringa(1992)}]{Wieringa1992}
Wieringa, M.~H. 1992, Experimental Astronomy, 2, 203,
  \dodoi{10.1007/BF00420576}

\bibitem[{Yatawatta(2015)}]{Yatawatta2015}
Yatawatta, S. 2015, Monthly Notices of the Royal Astronomical Society, 449,
  4506, \dodoi{10.1093/mnras/stv596}

\bibitem[{Yatawatta {et~al.}(2009)Yatawatta, Zaroubi, {De Bruyn}, Koopmans, \&
  Noordam}]{Yatawatta2009}
Yatawatta, S., Zaroubi, S., {De Bruyn}, G., Koopmans, L., \& Noordam, J. 2009,
  in 2009 IEEE 13th Digital Signal Processing Workshop and 5th IEEE Signal
  Processing Education Workshop, DSP/SPE 2009, Proceedings, 150--155,
  \dodoi{10.1109/DSP.2009.4785912}

\bibitem[{Yatawatta {et~al.}(2013)Yatawatta, {De Bruyn}, Brentjens,
  Labropoulos, Pandey, Kazemi, Zaroubi, Koopmans, Offringa, Jeli{\'{c}},
  {Martinez Rubi}, Veligatla, Wijnholds, Brouw, Bernardi, Ciardi, Daiboo,
  Harker, Mellema, Schaye, Thomas, Vedantham, Chapman, Abdalla, Alexov,
  Anderson, Avruch, Batejat, Bell, Bell, Bentum, Best, Bonafede, Bregman,
  Breitling, {Van De Brink}, Broderick, Br{\"{u}}ggen, Conway, {De Gasperin},
  {De Geus}, Duscha, Falcke, Fallows, Ferrari, Frieswijk, Garrett, Griessmeier,
  Gunst, Hassall, Hessels, Hoeft, Iacobelli, Juette, Karastergiou, Kondratiev,
  Kramer, Kuniyoshi, Kuper, {Van Leeuwen}, Maat, Mann, McKean, Mevius, Mol,
  Munk, Nijboer, Noordam, Norden, Orru, Paas, Pandey-Pommier, Pizzo, Polatidis,
  Reich, R{\"{o}}ttgering, Sluman, Smirnov, Stappers, Steinmetz, Tagger, Tang,
  Tasse, {Ter Veen}, Vermeulen, {Van Weeren}, Wise, Wucknitz, \&
  Zarka}]{Yatawatta2013a}
Yatawatta, S., {De Bruyn}, A.~G., Brentjens, M.~A., {et~al.} 2013, Astronomy
  and Astrophysics, 550, A136, \dodoi{10.1051/0004-6361/201220874}

\bibitem[{Zhang {et~al.}(2020)Zhang, Pober, Li, Hazelton, Morales, Trott,
  Jordan, Joseph, Beardsley, Barry, Byrne, Tingay, Chokshi, Hasegawa, Jacobs,
  Lanman, Line, Lynch, McKinley, Mitchell, Murray, Pindor, Rahimi, Takahashi,
  Wayth, Webster, Wilensky, Yoshiura, \& Zheng}]{Zhang2020}
Zhang, Z., Pober, J.~C., Li, W., {et~al.} 2020, Publications of the
  Astronomical Society of Australia, 37, E045, \dodoi{10.1017/pasa.2020.37}

\end{thebibliography}
\bibliographystyle{aasjournal}



\end{document}